\documentclass[lettersize,journal]{IEEEtran}

\usepackage{graphicx}           
\usepackage{amsmath,amsfonts}   
\usepackage{multirow}           
\usepackage{algorithmic}       
\usepackage{textcomp}           

\usepackage{booktabs}           
\usepackage{adjustbox}          
\usepackage{float}              

\usepackage[table]{xcolor}      
\definecolor{lightgreen}{rgb}{0.85, 1.0, 0.85}

\usepackage{caption}            
\usepackage{subcaption}         

\usepackage{tcolorbox}          
\usepackage{hyperref}

\usepackage[linesnumbered,algoruled,boxed,lined]{algorithm2e}

\usepackage{subcaption}
\usepackage[utf8]{inputenc}
\usepackage{flushend}
\SetKwRepeat{Do}{do}{while}
\usepackage{orcidlink}
\let\oldnl\nl
\newcommand{\nonl}{\renewcommand{\nl}{\let\nl\oldnl}}
\usepackage{float}
\newtcolorbox{custombox}[1]{
	colback=gray!10,
	colframe=gray!20,
	left=1mm,
	right=1mm,
	top=1mm,
	bottom=1mm,
	fonttitle=\bfseries,
	arc=2mm,
	leftrule=0mm,
	rightrule=.5mm,
	toprule=0mm,
	bottomrule=.5mm,
	notitle,
	before=\par\smallskip\noindent,
	before upper={\textbf{#1: } },
}

\usepackage{amssymb}
\newsavebox\CBox

\newcounter{finding}
\newcommand\pimodel{$\pi_0$~} 
\newtcolorbox{FindingBox}[1][]{%
  colframe=black!50,
  colback=white,
  sharp corners,
  left=4pt,
  right=4pt,
  top=2pt,
  bottom=2pt,
  boxsep=2pt,
  before skip=6pt,
  after skip=6pt,
  fontupper=\small,
  before upper={%
    \stepcounter{finding}%
    \textbf{Finding~\thefinding. }%
  },
  #1
}

\newcommand{\revision}[1]{{\color{black}#1}}

\begin{document}

\title{Evaluating Uncertainty and Quality of Vision-Language-Action-enabled Robots}

\author{Pablo~Valle~\orcidlink{0000-0002-0588-316X}, Chengjie~Lu~\orcidlink{0000-0002-5818-7547}, Shaukat~Ali~\orcidlink{0000-0002-9979-3519} and 
Aitor~Arrieta~\orcidlink{0000-0001-7507-5080}%
\thanks{Pablo Valle and Aitor Arrieta are with Mondragon University, Guipuzcoa, Spain. E-mail: \url{pvalle@mondragon.edu}, \url{aarrieta@mondragon.edu}. Chengjie Lu and Shaukat Ali are with Simula Research Laboratory, Oslo, Norway. E-mail: \url{chengjielu@simula.no}, \url{shaukat@simula.no}).}%
}
       
\markboth{}%
{Shell \MakeLowercase{\textit{et al.}}: A Sample Article Using IEEEtran.cls for IEEE Journals}

\maketitle

\begin{abstract}
Vision-Language-Action (VLA)-enabled robots integrate visual perception, natural language understanding, and action planning to interpret their environment, comprehend instructions, and perform embodied tasks autonomously. 
Such robots are typically evaluated through task success rates, \revision{i.e., whether a robot performs its intended task, which are commonly used as test oracles for evaluating such robots.} Such an evaluation fails to capture the quality of task execution and the robot’s confidence in its decisions. In this paper, we \revision{adapt} eight uncertainty metrics and five quality metrics \revision{specifically designed for VLA-enabled robotic manipulation tasks}. We assess their effectiveness through a large-scale empirical study involving 908 successful task executions from three state-of-the-art VLA models across four representative robotic manipulation tasks \revision{and two robot embodiments}. Human domain experts manually labeled task quality, enabling us to analyze the correlation between our proposed metrics and expert judgments, \revision{serving as a human oracle for testing such robots.} The results reveal that several metrics show moderate to strong correlation with human assessments, highlighting their utility for evaluating task quality and model confidence. Furthermore, we found that some metrics can discriminate between high-, medium-, and low-quality executions from unsuccessful tasks, which is useful when test oracles are absent. Our findings challenge the adequacy of current evaluation practices that rely solely on binary success rates and pave the way for improved real-time monitoring and adaptive enhancement of VLA-enabled robots.

\end{abstract}


\section{Introduction}

Robotic systems are an integral part of today's manufacturing processes. Yet, their deployment is highly constrained and limited to businesses with high revenue due to the lack of qualified workers with robotic programming skills~\cite{Tsarouchi02082016}. Recent advances in generative artificial intelligence (AI) are reducing this gap~\cite{SOORI202354,wang2024large}. Companies like NVIDIA predict that these robots will soon be deployed in multiple areas beyond manufacturing, including apartments, offices, hotels, etc. Examples include Tesla's Optimus humanoid robot, Boston Dynamics' Spot robot, and Agibot's humanoid robots.

The main generative AI-based state-of-the-art models that enable this are Vision-Language-Action (VLA) models, \revision{which are typically integrated within complete robotic systems. These systems combine VLA models with additional robotic modules such as low-level motion controllers, actuation systems, and safety or stabilization controllers~\cite{kim2024openvla,zitkovich2023rt}}. VLA-enabled robots represent a promising approach to enabling robots to interpret visual scenes, understand natural language commands, and execute complex tasks seamlessly in dynamic environments. These \revision{systems} take as input a set of images, a natural language instruction, and the state of the robot under control. As output, they generate a set of action chunks that are directly converted into loco-motion and manipulation commands. This effectively bridges high-level cognitive instructions with low-level robotic actions, simplifying robotic programming and facilitating human-robot interaction.

Different VLA models for robotic control have been recently proposed, including GR00T-N1~\cite{nvidia2025gr00tn1openfoundation}, \pimodel~\cite{black2024pi0visionlanguageactionflowmodel}, OpenVLA~\cite{kim2024openvla}, and SpatialVLA~\cite{qu2025spatialvla}. However, the assessment of these systems lacks standardization, as each developer typically proposes their own evaluation benchmarks due to the absence of a universally accepted benchmark. These benchmarks often involve multiple scenarios, each comprising a predefined set of objects and various instructions (e.g., \textit{``grasp a coke can''}). Given the non-systematic nature of these evaluations, recently, Wang et al.~\cite{wang2025vlatest} proposed VLATest, \revision{a fuzzing-based benchmarking framework for VLA-enabled robots}. Similar to studies proposing new VLA models, VLATest~\cite{wang2025vlatest} determines task success using symbolic oracles \revision{that assess only task completion by checking final states of the environment (e.g., whether an object reaches a target position).}

\revision{These oracles are limited, as they primarily capture functional correctness while ignoring other critical dimensions of system behavior. In particular, they do not assess the quality of task execution, including safety, smoothness, or efficiency of the full robotic software system. This indicates a key test-oracle problem in VLA-enabled robots, where correctness is reduced to binary success while rich behavioral properties (i.e., task quality) remain unobserved. Our analysis of successful executions from VLATest~\cite{wang2025vlatest} highlights this limitation: even when tasks are labeled as successful, many executions exhibit poor-quality behavior. For instance, robots may collide with objects, follow inefficient or unstable trajectories, or require multiple attempts to complete simple actions due to repeated drops of different objects. In some cases, task success appears to result from chance rather than consistent robot performance, indicating weak behavioral correctness even when a test passes. These observations demonstrate that current test oracles are insufficient to assess the overall performance of VLA-enabled robots and provide limited assurance regarding real-world reliability.}


To address these issues, we \revision{adapt} and investigate five quality metrics to assess whether a task is performed in a high-quality manner. In addition, we \revision{adapt} eight uncertainty metrics to quantify the confidence \revision{of the robot} during task execution. The underlying intuition is that higher uncertainty corresponds to lower confidence, which often results in low-quality task execution. \revision{In addition, these metrics are applicable both at design time, serving as an oracle for evaluating robot behavior during development, and at runtime, where they can be used for monitoring, anomaly detection, and early identification of potential failures. Furthermore, they can support falsification-based test generation and serve as adequacy criteria for robotic test suites}.

Specifically, the key contributions of this paper can be summarized as follows: 

\begin{itemize}
    \item \revision{We design an uncertainty and quality evaluation framework for VLA-enabled robots, which integrates eight uncertainty and five quality metrics. 
    }
    \item \revision{We propose a set of quality evaluation guidelines for human experts to evaluate the quality of VLA-enabled robotic tasks execution.}
    \item \revision{We conduct an extensive empirical study with three state-of-the-art VLA models integrated in the robots across four tasks. Besides, three domain experts manually analyze and label the quality of 908 successful test executions, and then we evaluate the proposed quality and uncertainty metrics by measuring their correlation with these annotations.}
    \item We provide a complete replication package~\cite{pvalle_VLA-UQ}, including code, configuration files, and instructions to facilitate reproducibility and further research. In addition, we provide a Zenodo package~\cite{valle2025vla} including all the results from our experiments.

\end{itemize}

The results and findings of our study reveal critical shortcomings in current assessment approaches for \revision{VLA-enabled robotic systems}. Although traditional evaluations have primarily relied on binary task completion success, our thorough analysis of 908 successful executions across three prominent VLA models uncovered striking differences in task execution quality. For instance, the \pimodel{} model exhibited poor execution quality, a significant finding that starkly contrasts with their generally optimistic self-assessments. Our domain experts' labeling of successful tasks identified pronounced disparities, underscoring the inadequacy of success rate alone as a reliable evaluation metric. Furthermore, \revision{the adapted} eight uncertainty metrics and five quality metrics revealed moderate to high correlation with expert evaluations, suggesting their effectiveness in capturing model performance discrepancies. These findings not only challenge existing evaluation frameworks but also suggest essential future avenues toward improved real-time monitoring and adaptive enhancements in robotic systems.

\section{\textcolor{black}{Background on Vision-Language-Action-Enabled Robots}}
\label{sec:Back_VLA}

\revision{Recent advances in robotics have been driven by the integration of multimodal learning systems into robotic control. In particular}, Vision-Language-Action (VLA)  models~\cite{nvidia2025gr00tn1openfoundation,black2024pi0visionlanguageactionflowmodel,kim2024openvla,qu2025spatialvla} \revision{have emerged as key systems to enable robots to interpret complex environments and execute high-level instructions. This enables robots as embodied agents that combine perception, reasoning, and control to interact with the physical world.}

\revision{Traditional robotic systems rely on modular pipelines, where perception, planning, and control are designed independently. In such traditional architectures}, Deep Neural Networks (DNN) are typically designed for single-domain tasks. For instance, convolutional architectures excel at extracting hierarchical features from images \cite{lecun2002gradient,krizhevsky2012imagenet,simonyan2014very,szegedy2015going}, while transformer-based \cite{vaswani2017attention} language models capture long-range dependencies in text \cite{achiam2023gpt,Claude:online,team2023gemini,grattafiori2024llama,jiang2023mistral7b}. In contrast, VLA models encode visual observations (e.g., images or video frames) and textual instructions into a shared embedding space, from which they generate structured sequences of actions. This approach allows VLA models to interpret instructions such as ``\textit{Pick up the Apple}'' in the context of a scene, where they must resolve ambiguities in object references using visual attention mechanisms and translate the instruction into a precise sequence of physical movements.

\begin{figure*}[ht]
    \centering
    \includegraphics[width=\textwidth]{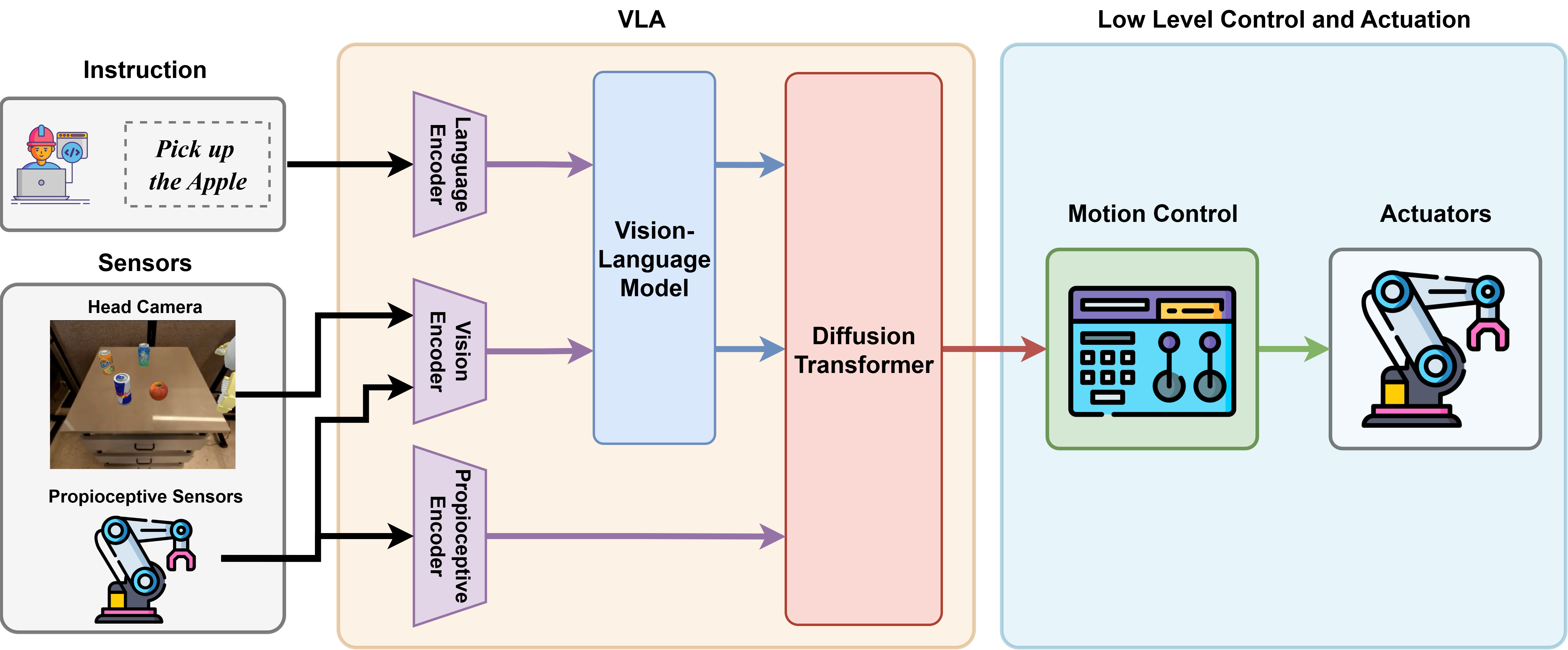}
    \caption{\revision{Overview of the Architecture of a VLA-Enabled Robot}}
    \label{fig:VLA-arch}
\end{figure*}

As Figure~\ref{fig:VLA-arch} depicts,\revision{VLA-enabled robots comprise three main modules and a user-provided instruction. First, the Sensors module reafs information from the environment, including visual observations captured by the head camera and the robot's internal state obtained through proprioceptive sensors. These observations, together with the user instruction, are provided as input to the VLA module, which is responsible for generating the actions required to accomplish the specified task. Within the VLA module, dedicated encoders process visual observations, proprioceptive data, and natural language instructions to produce representations that can be jointly interpreted by the Vision-Language Model. Based on these representations, a Diffusion Transformer generates the sequence of actions to be executed by the robot. The generated actions are then passed to the Low-Level Control and Actuation module, which translates them into actuator commands and executes them on the robot. The motion controller ensures that the generated commands satisfy platform-specific constraints, such as velocity limits and joint range restrictions, before forwarding them to the actuators.By having a closer look to the controller (i.e., the VLA model) of a VLA-enabled robot, we can see that it operates through an integrated pipeline} encompassing perception, reasoning, and control. Given an observation $o_t = \left[I^t_1,\, I^t_2,\, \dots,\, I^t_n,\, \ell_t,\, q_t\right]$, which includes $n$ RGB images ($I^t_n$), a language instruction ($\ell_t$), and the robot's proprioceptive state ($q_t$), the visual encoder first processes the image into feature maps that capture both local structure and global context, while the language encoder converts the instruction into a dense semantic embedding. Simultaneously, the robot's proprioceptive state is passed through an encoder that extracts relevant internal state features. These embeddings are projected into a shared latent space, aligning linguistic tokens with relevant visual regions and robotic states. From this mixed representation, the action decoder outputs an action chunk $A_t = \left[a_t,\, a_{t+1},\, \dots,\, a_{t+H-1}\right]$, where each $a_{t+h}$ is a multidimensional control signal defined as $a_{t+h} = \left[a_{t+h,1},\, a_{t+h,2},\, \dots,\, a_{t+h,D}\right]$, with $D$ denoting the total number of control dimensions. Specifically, in most VLA models this is a seven-dimensional control signal: three dimensions for position ($x, y, z$), three dimensions for orientation ($roll, pitch, yaw$), and one dimension for opening of the gripper. Note that the action chunk size $H$, called action horizon, is variable across models; for instance, the action horizon of the \pimodel{} model is up to 50 actions~\cite{black2024pi0visionlanguageactionflowmodel}.

Recent VLA models~\cite{nvidia2025gr00tn1openfoundation, black2024pi0visionlanguageactionflowmodel} introduced a diffusion-based denoising process before providing the actions. In these types of models, the action chunk $A_t$ output by the VLA model is refined through a denoising diffusion process. This process involves iteratively removing noise from the initial action prediction using contextual information extracted from the representation of the visual input, language instruction, and proprioceptive state (i.e., $o_t$). Rather than sampling entire trajectories from scratch, the diffusion module acts as a post-processing step that enhances the base model outputs by steering them to more plausible regions of the action space. This approach has been shown to improve generalization in complex scenes and to increase robustness by reducing error accumulation in longer-horizon tasks~\cite{black2024pi0visionlanguageactionflowmodel,zhao2023learning, nvidia2025gr00tn1openfoundation}.

Training VLA models typically follows one of these two approaches: (1) training from scratch or (2) leveraging large-scale pretraining followed by targeted fine-tuning. In the former case, the model is initialized and trained on a dataset tailored to a specific robot or task environment, ensuring that the learned policies are aligned with the unique constraints of such a system \cite{team2024octo,brohan2022rt}. In the latter, which is an approach that scales better, the VLA model is pretrained on extensive multimodal datasets, such as the Open X-Embodiment datasets~\cite{o2024open}, comprising diverse language commands, visual observations, and action sequences from various domains. This pretraining phase provides the model with transferable representations and general reasoning capabilities. Subsequently, the model is fine-tuned on robot-specific datasets to adapt its action decoder to the unique constraints of the hardware (e.g., kinematics or joint limits). This fine-tuning is critical to ensure that the predicted actions are not only meaningful but also executable and safe within the physical constraints of the robot. 


\section{Uncertainty Metrics}\label{sec:UncertaintyMetrics}

We \revision{adopt} a total of eight uncertainty metrics for VLA-enabled robots \revision{in robotic software}, aimed at quantifying model confidence. Our selection is guided by the need to capture different aspects of uncertainty in multimodal tasks. \revision{The selected metrics were chosen to reflect the two fundamental components of VLA architectures: (i) probabilistic action generation, where uncertainty is expressed through token probabilities, and (ii) embodied action execution, where uncertainty may manifest as unstable or inconsistent control behaviors. Together, these perspectives allow uncertainty to be characterized both at the model prediction level and at the action generation level.} Specifically, we adapt four confidence-based metrics commonly used in deep learning models, such as Max Probability~\cite{10.1145/3417330}, Prediction Confidence Score (PCS)~\cite{scheffer2001active}, Entropy~\cite{shannon1948mathematical}, and DeepGini~\cite{feng2020deepgini}. \revision{In the context of VLA-enabled robots, we compute them over token output probabilities of the VLA model, referred to as token-based metrics. While several VLA policies may generate continuous actions (e.g., diffusion-based policies), in those cases we extract the discrete token representation from the model prior to the diffusion process (as detailed in Appendix~\ref{app:ModelArch}), which provides a shared vocabulary over which probabilities are defined, enabling a unified token-based formulation across all evaluated models.}
These were chosen because they directly reflect prediction confidence and have interpretable probabilistic principles that remain relevant when extended to the outputs of VLA models. In addition to these, since the output of VLA models involves structured decision sequences, we \revision{use four additional} metrics specifically \revision{adapted} to measure the uncertainty in generated actions. These metrics, different but complementary to more traditional token-based metrics, aim to capture subtle differences in robot action variability and model agreement. By combining both \revision{types of} metrics, we aim to offer a broad set of metrics for evaluating uncertainty in VLA models. 

\subsection{\revision{Token-based Uncertainty Metrics}}

\revision{We adopt four different token-based uncertainty metrics, originally defined for Large Language Models (LLMs)~\cite{huang2023look}, and adapt them to evaluate uncertainty in VLA-enabled robots. In these models, actions are represented as sequences of discrete tokens, each associated with a probability distribution over possible outputs. At each time step $t$, the model predicts an action composed of \revision{a set of tokens $TN$}, where each token \revision{$a_{t,tn} \in TN$} is associated with a probability distribution $\{p_i\}$. We compute uncertainty at the token level and aggregate it across all tokens to obtain a step-level uncertainty score $u_t$. Higher values of $u_t$ consistently indicate higher uncertainty. For diffusion-based VLA models, token probabilities are obtained from the pre-diffusion discrete representation described above, ensuring compatibility with all evaluated architectures. Sequence-level temporal dependence is handled implicitly via the autoregressive decoding process, while uncertainty is computed per step rather than over full trajectories. We note that token-based metrics are designed to measure uncertainty in the learned policy distribution (i.e., model-internal uncertainty) and are therefore primarily driven by decoding stochasticity, rather than external environmental noise.}

\subsubsection{\revision{Token Probability (TB-TP)}}

\revision{This metric captures uncertainty based on the maximum predicted probability for each token. It reflects how strongly the model prefers its top prediction. When the model assigns a high probability to a single token, it indicates confident and decisive behavior. Conversely, lower maximum probabilities suggest that the model is uncertain and distributes probability across multiple alternatives. TB-TP is an adaptation of standard confidence scoring from probabilistic classification and language modeling, applied here to discretized action tokens rather than continuous control outputs. This metric can be formalized as follows:}

\begin{equation}
    u_t = 1 - \frac{1}{TN} \sum_{tn=1}^{TN} \max_i \, p_i(a_{t,tn})
\end{equation}

\subsubsection{\revision{Prediction Confidence Score (TB-PCS)}}

\revision{TB-PCS measures the margin between the two most probable predictions for each token. This metric captures local ambiguity, even if the top prediction has a relatively high probability, a small gap with the second-best alternative indicates uncertainty in distinguishing between competing actions. In contrast, a large margin reflects clear preference and higher confidence. This metric can be formalized as follows:}

\begin{equation}
    PCS(a_{t,tn}) = \max_i \, p_i - \textit{second-max}_i \, p_i
\end{equation}
\begin{equation}
    u_t = 1 - \frac{1}{TN} \sum_{tn=1}^{TN} PCS(a_{t,tn})
\end{equation}

\subsubsection{DeepGini (TB-D)}

\revision{DeepGini evaluates the overall concentration of the predicted probability distribution. It considers all possible token probabilities rather than focusing only on the top predictions. A highly peaked distribution (i.e., one dominant token) results in low uncertainty, while a flatter distribution, where multiple tokens have comparable probabilities, indicates higher uncertainty.We adopt DeepGini from prior work in neural network testing~\cite{feng2020deepgini} and adapt it to token-level action distributions in VLA models. Its use is justified by its ability to capture global dispersion in the action probability distribution, which complements max-probability and entropy-based measures. This metric can be formalized as follows:}

\begin{equation}
    u_t = \frac{1}{TN} \sum_{tn=1}^{TN} \left( 1 - \sum_{i=1}^{n} p_i^2(a_{t,tn}) \right)
\end{equation}

\subsubsection{Entropy (TB-E)}

\revision{Entropy measures the uncertainty by quantifying the randomness of the predicted distribution. Similarly to Deepgini, entropy uses the entire distribution and penalizes both uniformity and spread. Higher entropy values indicate that the model is less certain about its prediction, as probability is distributed across many possible tokens. This metric can be formalized as follows:}

\begin{equation}
    u_t = \frac{1}{TN} \sum_{tn=1}^{TN} \left( - \sum_{i=1}^{n} p_i(a_{t,tn}) \log p_i(a_{t,tn}) \right)
\end{equation}

\subsection{\revision{Action-based Uncertainty Metrics}}

\revision{While token-based metrics quantify uncertainty directly from the probability distributions produced by the VLA model, they do not capture how uncertainty is reflected in the generated robot behavior. VLA-enabled robots operate through a perception-action loop that integrates visual observations, language instructions, and robot states to generate control actions. When the model encounters ambiguous observations, underspecified instructions, or situations that are poorly represented in the training data, uncertainty may manifest as inconsistent or unstable action sequences. Such behaviors are commonly associated with unreliable policy execution in robotics, where temporally coherent control signals generally correspond to more robust and predictable behavior.}

\revision{To complement token-based uncertainty metrics, we introduce action-based uncertainty metrics that operate directly on the continuous action vectors generated by the policy. These metrics quantify temporal consistency in the action sequence through successive finite differences, providing a behavioral perspective on uncertainty. The underlying intuition is that uncertainty may emerge at different temporal scales. First-order differences capture abrupt changes between consecutive actions, second-order differences capture variations in action velocity, and third-order differences capture higher-frequency fluctuations in the action dynamics. Similar formulations have been widely used in robotics and control to characterize trajectory smoothness and stability~\cite{hogan2009sensitivity, PHANG201565}. In this work, we adapt them as indicators of uncertainty in VLA-generated actions, as they enable the detection of oscillatory or inconsistent behaviors that may not be apparent from token-based metrics alone. Given a sequence of predicted actions ${a_1, a_2, \dots, a_T}$, where each action $a_t \in \mathbb{R}^D$ corresponds to a $D$-dimensional control signal, we define the following metrics.}

\subsubsection{\revision{Action Position Instability (A-PI)}}

\revision{A-PI measures the magnitude of changes between consecutive actions, capturing the local temporal consistency of the generated control sequence. In VLA-enabled robots, uncertainty may manifest as hesitation between competing actions, leading to abrupt changes in consecutive commands. Therefore, large action differences can indicate that the model is struggling to maintain a consistent behavioral strategy. This metric can be formalized as follows:}
\begin{equation}
\Delta a_t = a_t - a_{t-1}
\end{equation}
\begin{equation}
u_t = \frac{1}{D} \sum_{d=1}^{D} \left| \Delta a_{t,d} \right|
\end{equation}

\revision{Lower values correspond to stable and temporally coherent action sequences, while higher values indicate abrupt behavioral changes that may be associated with increased uncertainty or ambiguity during decision making.}

\subsubsection{\revision{Action Velocity Instability (A-VI)}}

\revision{A-PI may overestimate uncertainty in scenarios where the robot moves consistently at high speed. To distinguish between purposeful motion and instability, A-VI measures changes in velocity through the second-order difference of the action sequence. Rather than quantifying the amount of motion, it captures variations in motion dynamics. Consequently, A-VI is sensitive to oscillatory behavior and repeated corrections, which are common manifestations of uncertainty when the model alternates between different possible actions. This metric can be formalized as follows:}
\begin{equation}
\Delta^2 a_t = a_t - 2a_{t-1} + a_{t-2}
\end{equation}
\begin{equation}
u_t = \frac{1}{D} \sum_{d=1}^{D} \left| \Delta^2 a_{t,d} \right|
\end{equation}

\revision{Higher values indicate abrupt changes in velocity, revealing behavioral oscillations and instability that may not be apparent from position differences alone.}

\subsubsection{\revision{Action Acceleration Instability (A-AI)}}\label{sec:A-VI}

\revision{To further capture fine-grained temporal inconsistencies, A-AI measures the rate of change of acceleration through the third-order difference. While A-PI and A-VI detect abrupt positional and velocity variations, respectively, they may not capture small but persistent oscillations that accumulate over time. A-AI is particularly sensitive to repeated micro-corrections and high-frequency fluctuations in the generated actions, which frequently arise when the model alternates between multiple plausible control decisions. In high-dimensional action spaces, these effects may be difficult to identify from individual action dimensions; therefore, we compute the metric independently for each action dimension and aggregate the results, allowing localized instability patterns to contribute to the overall uncertainty estimate. This metric can be formalized as follows:}
\begin{equation}
\Delta^3 a_t = a_t - 3a_{t-1} + 3a_{t-2} - a_{t-3}
\end{equation}
\begin{equation}
u_t = \frac{1}{D} \sum_{d=1}^{D} \left| \Delta^3 a_{t,d} \right|
\end{equation}

\revision{Higher A-AI values indicate less smooth and more erratic action evolution, suggesting increased instability in the generated control policy.}

\subsubsection{\revision{Execution Variability (EV)}}\label{sec:EV}

\revision{Execution Variability captures uncertainty by measuring the consistency of the predicted actions across repeated inferences using identical inputs. Since the observation and instruction remain fixed, variations across repeated runs primarily arise from the model itself and the stochastic decoding process. Consequently, EV serves as a proxy for epistemic uncertainty, reflecting uncertainty in the learned policy rather than variability introduced by changes in the environment. This metric can be formalized as follows:}
\begin{equation}
u_t = \frac{1}{D} \sum_{d=1}^{D}
\sqrt{
\frac{1}{N} \sum_{n=1}^{N}
\left(a_{t,d,n} - \frac{1}{N} \sum_{m=1}^{N} a_{t,d,m} \right)^2
}
\end{equation}

\noindent \revision{where $N$ is the number of repeated inferences.}

\revision{As EV quantifies uncertainty by measuring the dispersion of the generated action vectors across repeated inferences for the same input, lower EV values indicate that the policy consistently produces similar actions, reflecting high confidence in the selected behavior. Conversely, high EV values indicate greater disagreement among the generated actions and therefore higher uncertainty. Unlike token-based metrics, which estimate uncertainty from probability distributions, EV characterizes uncertainty directly from the variability of the resulting control actions.}

\section{Quality Metrics}\label{sec:QualityMetrics}

In this section, we introduce a set of \revision{trajectory-based quality metrics adapted from classical robotic smoothness and motion quality measures} to assess the quality of task execution in VLA-enabled robots. Unlike the uncertainty metrics presented in the previous section, which quantify the confidence and consistency of the actions generated from the VLA model, these metrics evaluate the characteristics of the executed robot behavior and its progression toward the task objective. Consequently, uncertainty metrics primarily characterize the policy generation process, whereas quality metrics characterize the resulting trajectory and task execution. \revision{The proposed metrics are adapted formulations of established trajectory quality criteria for VLA-enabled robots, including acceleration- and jerk-based measures commonly used in robotics. To the best of our knowledge, this is the first work to systematically adopt and evaluate such metrics in the context of VLA-enabled robots. These metrics aim to capture properties of the executed behavior that are not reflected by task success alone, including smoothness, stability, efficiency, and human-perceived motion quality. Together, they provide a more comprehensive characterization of task execution quality and enable a deeper understanding of how naturally and reliably VLA-enabled robots interact with their environment.}

\subsection{Trajectory Position Instability (TCP-PI)}

TCP Trajectory Position Instability (TCP-PI) \revision{measures the magnitude of positional changes of the Tool Center Point (TCP) between consecutive time steps, capturing the smoothness of the executed trajectory. Large variations may indicate abrupt movements or inconsistent motion. This metric is closely related to classical first-order finite-difference smoothness measures used in robotics and trajectory optimization. It can be defined as follows: }
\begin{equation}
\Delta p_t = p_t - p_{t-1},
\end{equation}
\begin{equation}
\text{q}_t = \left| \Delta p_t \right|
\end{equation}

\revision{\noindent where $p_t$ is the position of teh TCP defined as $p_t= [x_t, y_t, z_t]$. All values are computed in task (Cartesian) space and aggregated over time using a mean over the trajectory. To ensure comparability across executions, all trajectories are resampled to a fixed temporal resolution, reducing sensitivity to trajectory length and sampling frequency. Lower values correspond to smoother and more stable trajectories, while higher values reflect rapid or irregular movements.}

\subsection{Trajectory Velocity Instability (TCP-VI)}

As discussed in Section~\ref{sec:A-VI}, large positional changes do not necessarily indicate erratic motion. In some cases, they can result from high-speed movements. \revision{To address this, we introduce Trajectory Velocity Instability (TCP-VI), which measures second-order differences and is closely related to acceleration-variation and jerk-based smoothness measures commonly used in robotics (e.g., minimum-jerk trajectory formulations). It can be defined as}:
\begin{equation}
\Delta^2 p_t = \Delta p_t - \Delta p_{t-1} = p_t - 2p_{t-1} + p_{t-2}
\end{equation}
\begin{equation}
\text{q}_t = \left| \Delta^2 p_t \right|
\end{equation}

\revision{This metric captures variations in motion dynamics, where lower values indicate consistent velocity profiles, whereas higher values indicate irregular acceleration patterns. We compute this metric in task space using uniformly resampled trajectories with fixed timestep resolution to ensure invariance to trajectory duration and sampling rate differences.}

\subsection{Trajectory Acceleration Instability (TCP-AI)}

\revision{TCP-AI further refines the analysis of previous metrics by measuring the rate of change of acceleration, making it sensitive to high-frequency oscillations. This metric is closely related to jerk-based smoothness measures widely used in motion planning and human movement analysis and is defined as follows}
\begin{equation}
\Delta^3 p_t = \Delta^2 p_t - \Delta^2 p_{t-1} = p_t - 3p_{t-1} + 3p_{t-2} - p_{t-3}
\end{equation}
\begin{equation}
\text{q}_t = \left| \Delta^3 p_t \right|
\end{equation}
\noindent where \revision{higher values indicate abrupt transitions in acceleration, often perceived as jittery or unstable motion resulting in lower perceived quality for the user. This formulation corresponds to a discrete approximation of jerk and is sensitive to sampling frequency; therefore, all trajectories are normalized to a fixed timestep resolution before computation. All metrics are computed in task space, which aligns with human perceptual evaluation of motion quality in robotic execution.} This anti-pattern was already identified as an issue in the GR00T-N1 repository from practitioners~\footnote{\url{https://github.com/NVIDIA/Isaac-GR00T/issues/114}}.

\subsection{Trajectory Instability (TI)}

The Trajectory Instability (TI) quality metric aims to quantify the quality of the task by analyzing the physical behavior of the robot, specifically the evolution of its end-effector motion over time. \revision{While previous metrics operate at the step level, Trajectory Instability (TI) evaluates the overall smoothness of the trajectory. This metric is based on the jerk, the third derivative of position, which has been widely used to characterize motion quality in motor control and UAV control domains~\cite{hogan2009sensitivity, PHANG201565}.} 
In addition, the study performed by Flash et al.~\cite{flash1985coordination}, showed that human movements follow a minimum-jerk principle. From this perspective, movements with minimal jerk are considered the smoothest, as lower jerk reflects gradual, continuous changes in acceleration without abrupt transitions.

\revision{Let $\vec{r}(t) = (x(t), y(t), z(t))$ denote the trajectory. The jerk $\vec{j}(t)$ is defined as the third derivative of position. We compute its magnitude at each time step:}
\begin{equation}
\lVert \vec{j}_t \rVert = \sqrt{j_{x,t}^2 + j_{y,t}^2 + j_{z,t}^2}
\end{equation}

To aggregate this information over the entire trajectory as a quality indicator, we compute the Root Mean Square (RMS) jerk, as proposed by Hogan et al.~\cite{hogan2009sensitivity}, which serves as an overall measure of the motion quality across all poses during task execution:
\begin{equation}
q = \sqrt{\frac{1}{T - 3} \sum_{t=1}^{T - 3} \lVert \vec{j}_t \rVert^2}
\end{equation}






\noindent where $T$ is the total number of samples. This provides a robust measure of motion smoothness: \revision{lower values indicate smoother, more consistent trajectories, while higher values correspond to oscillatory or abrupt motion patterns.}

\subsection{Optimal Trajectory Difference (OT)}
The Optimal Trajectory Difference (OT) evaluates the quality of robotic task execution by measuring the spatial proximity between the robot’s end-effector and task-relevant reference positions. Unlike motion-centric metrics, which assess smoothness or stability (e.g., Trajectory Instability), this metric focuses on goal-oriented behavior. The core intuition is that an effective trajectory should exhibit a consistent reduction in distance to the goal object or position as the task progresses. Deviations from this expected pattern, such as increasing distance, can indicate hesitation or suboptimal decision making.

\revision{At each time step $t$, we define a distance metric $d_t$ that depends on the task phase. For manipulation tasks such as \textit{``Move''}, \textit{``Put In''}, and \textit{``Put On''}, the reference dynamically changes depending on whether the object has been grasped. Before grasping, the metric encourages the robot to approach the object while also considering its relation to the final target. After grasping, the focus shifts exclusively to reaching the final placement location.}

\begin{equation}
\resizebox{\linewidth}{!}{$
d_t =
\begin{cases}
\left\| \vec{p}_{tcp}(t) - \vec{p}_{obj}(t) \right\| + \left\| \vec{p}_{tcp}(t) - \vec{p}_{end} \right\|, & \text{if object not grasped} \\
\left\| \vec{p}_{tcp}(t) - \vec{p}_{end} \right\|, & \text{if object grasped}
\end{cases}$}
\end{equation}

\revision{To ensure comparability across tasks and trajectories, the distance is normalized by the maximum achievable distance in the workspace, denoted as $d_{\max}$, which corresponds to the largest Euclidean distance between the target point and the furthest point.}
\begin{equation}
\tilde{d}_t = \frac{d_t}{d_{\max}}
\end{equation}

\revision{To evaluate behavioral consistency, we analyze the temporal evolution of the distance:}
\begin{equation}
\Delta \tilde{d}_t = \tilde{d}_t - \tilde{d}_{t-1}
\end{equation}

\revision{A negative value ($\Delta \tilde{d}_t < 0$) indicates progress toward the goal, while a positive value suggests divergence or inefficient behavior. We normalize this variation to obtain the final quality score:}
\begin{equation}
q_t = \frac{1}{2} (1 + \Delta \tilde{d}_t)
\end{equation}

\revision{Lower values of $q_t$ correspond to more effective and goal-consistent behavior, whereas higher values indicate suboptimal or unstable task execution.}

\section{Empirical Study}
\label{empirical_study}

We conducted an empirical study to investigate the performance of the proposed uncertainty and quality metrics for VLA-enabled Robotic Systems. This section presents our research questions (RQs) and describes the evaluation setup.

\subsection{Research Questions}
We evaluated the effectiveness and execution overhead of the uncertainty and quality metrics across three VLA models and four tasks used in the evaluation of VLATest~\cite{wang2025vlatest}. 
Specifically, we aim to answer the following RQs:
\begin{itemize}
    \item \textbf{RQ1 -- Replication:} \textit{To what extent do success tasks in VLATest align with human judgments of task quality?} We replicate the experiments conducted by Wang et al. \cite{wang2025vlatest} to assess the quality of the successful tasks.
    To this end, we conducted a more detailed analysis, using human experts, of the test cases that were classified as successful by VLATest~\cite{wang2025vlatest}.
    
    \item \textbf{RQ2 -- Correlation:} \textit{How accurately do the proposed uncertainty and quality metrics reflect the performance of the robot?} We assess whether the proposed uncertainty and quality metrics can be used as indicators of robot performance degradation. To this end, we examine the correlation of the proposed metrics in Sections~\ref{sec:UncertaintyMetrics} and~\ref{sec:QualityMetrics} with the quality level labeled by domain experts.

    \item \textbf{RQ3 -- Discrimination:} \textit{To what extent can the proposed metrics distinguish between successful and failing tasks?} We investigate whether the proposed metrics are associated with the robot's task success. By analyzing the distributions and effect sizes of each metric across tasks, we aim to assess their potential as indicators of task success or failure.
    
    \item \textbf{RQ4 -- Overhead:} \textit{How does integrating these metrics affect inference time?} Inference time of VLA models is critical as the control of robotic systems is carried out in real-time. Therefore, it is paramount to study the overhead produced by the different metrics to guide future practitioners in their adoption. This RQ studies the computational cost of each of the proposed uncertainty and quality metrics. 
\end{itemize}

\subsection{VLA Models}

In our evaluation, we used three state-of-the-art VLA models, OpenVLA~\cite{kim2024openvla}, SpatialVLA~\cite{qu2025spatialvla}, and \pimodel~\cite{black2024pi0visionlanguageactionflowmodel}. These three models were selected because (1) they are relatively recent, (2) they are relatively well-known, and (3) their performance reported in studies~\cite{kim2024openvla, black2024pi0visionlanguageactionflowmodel, qu2025spatialvla} was quite high. Moreover, these three models were available and fine-tuned for the robot of an open-source simulator. 
We used the fine-tuned versions adapted to two benchmark datasets, each corresponding to two of the task categories in our evaluation suite.The models for the \textit{Pick up} and \textit{Move Near} tasks were trained for the Google Robot, while those for the \textit{Put on} and \textit{Put in} tasks were trained for the WidowX robot. The specific model checkpoints used for each model are detailed below:
\begin{itemize}
    \item \textbf{OpenVLA:} We used the base checkpoint released by the authors on HuggingFace~\cite{openvla:online}. 
    \item \textbf{\pimodel:} We used the checkpoints provided by the authors of SpatialVLA, which were used in their evaluation. Two versions of the model were employed, one for each dataset. The checkpoints for the Fractal dataset are available at~\cite{HaomingFractal:online} and those for the Bridge dataset are available at~\cite{HaomingBridge:online}.
    \item \textbf{SpatialVLA:} We used the checkpoint of the model pretrained on the mixture data of both datasets~\cite{IPECCOMM11:online} provided in the paper of the model~\cite{qu2025spatialvla}.
\end{itemize}

An extended explanation of each model's architecture is provided in Appendix~\ref{app:ModelArch}. Note that SpatialVLA and \pimodel{} were not used in VLATest~\cite{wang2025vlatest}, and therefore our study provides new results of these two models.

\subsection{Environments}\label{sec:environments}

As in VLATest~\cite{wang2025vlatest}, we conducted our evaluation using the \textit{SimplerEnv} benchmark~\cite{li24simpler}, assessing the effectiveness of our metrics across two robotic platforms and four different tasks in total. The first robotic arm was the so-called Google robot, an Everyday Robot\footnote{\url{https://everydayrobots.ai/}}, using the Fractal dataset~\cite{brohan2022rt} for training the VLA models. The second robotic arm was the WidowX robot, and the Bridge V2 dataset~\cite{walke2023bridgedata} was used to train the VLA models. For the evaluation, we used the same four tasks selected in~\cite{wang2025vlatest}, two for one of the robots and two for the other one, explained below:

\begin{itemize}
    \item \textbf{Task 1: Pick Up an Object.}
    The VLA model must identify a target object and generate control signals to grasp and lift it. A successful completion requires the robot to grasp the correct object and lift it at least 0.02 meters for five consecutive frames. This task was assessed using the Google robot.
    
    \item \textbf{Task 2: Move Object A near Object B.}  
    The VLA model must locate source object A and generate control signals to move it near the target object B. The task is considered successful if object A is positioned within 0.05 meters or closer from object B. This task was assessed using the Google robot.
    
    \item \textbf{Task 3: Stack Object A on Object B.}  
    The VLA model must place object A stably on top of object B. Success is defined as object A remains balanced on top of object B without toppling. This task was assessed using the WidowX robot.
    
    \item \textbf{Task 4: Place Object A Inside Object B.}  
    The VLA model must generate control signals to place object A fully inside object B (e.g., \textit{``Place the apple into the basket''}). The task is considered successful if object A is entirely inside object B. This task was assessed using the WidowX robot.
\end{itemize} 

For each task, we used the first 500 scenes generated by Wang et al.~\cite{wang2025vlatest}. In these scenes, the target object(s) were randomly selected, accompanied by 0 to 3 confounding objects. The position and pose of each object were randomly assigned, following certain constraints as explained in~\cite{wang2025vlatest}. To avoid collision overlaps, a minimum distance of 0.15 meters was maintained between objects during placement. Regarding the environment, the default lighting and camera pose settings were used.

\subsection{Configurations}\label{sec:Configs}

To ensure reproducibility and enable fair comparisons with future work, we report all relevant parameter settings used in our experiments. A critical component for replicating our results is the set of VLA model weights, which is described in Section~\ref{sec:environments}.
In addition to the model checkpoints, we used a fixed seed
across all models to initialize the environments. This ensured that the initial scene state remained consistent for all evaluations. 

Regarding the uncertainty and quality metrics' configuration parameters, we configured the EV metric with 4 inference samples. We selected this number as it was the highest number of additional model instances we could load on our hardware. The metrics TCP-PI, TCP-VI, TCP-AI, A-PI, A-VI, and A-AI require temporal differences for their computation. Therefore, we defined a time window to include data from previous steps. Since the values at the edges of the time window are particularly sensitive, we found that the minimum range of 4 time steps, needed to compute A-AI and TCP-AI, did not yield sufficiently accurate results. Therefore, we increased the window size to 8 time steps to improve reliability. We selected this number based on empirical observation: smaller windows (e.g., 5–6 steps) led to less stable estimations, particularly for A-AI and TCP-AI, which are sensitive to short-term fluctuations. Increasing the window to 8 provided a better trade-off between temporal context and computational cost, providing more consistent and reliable metric values.

\subsection{Execution Platform and Runs}
To accelerate the execution of experiments, we distributed the workload across two execution platforms. On the one hand, the experiments for the \pimodel{} model were executed on a server with an AMD EPYC 7773X CPU and a NVIDIA RTX A6000 GPU. On the other hand, all the experiments for OpenVLA and SpatialVLA were executed on a server with an AMD EPYC 7763 CPU and two NVIDIA A100-SXM4-80GB GPUs. In both servers, the operating system was a 64-bit Ubuntu 20.04 LTS with Python 3.10 and CUDA 12.8.

Due to differences in internal architecture and weights, each model exhibited varying levels of resource consumption, which in turn affected the execution time of the scene. On average, each scene execution took approximately 270 seconds for SpatialVLA,  235 seconds for the \pimodel, and 210 seconds for OpenVLA. Given that each of the four tasks included 500 scenes, the total GPU time required was: 4 (tasks) $\times$ 500 (scenes) $\times$ (270 + 235 + 210) seconds = 1,430,000 seconds = 397 hours of GPU.

\subsection{Human Evaluation Procedure}

\begin{figure}[h]
    \centering
    \includegraphics[width=0.485\textwidth]{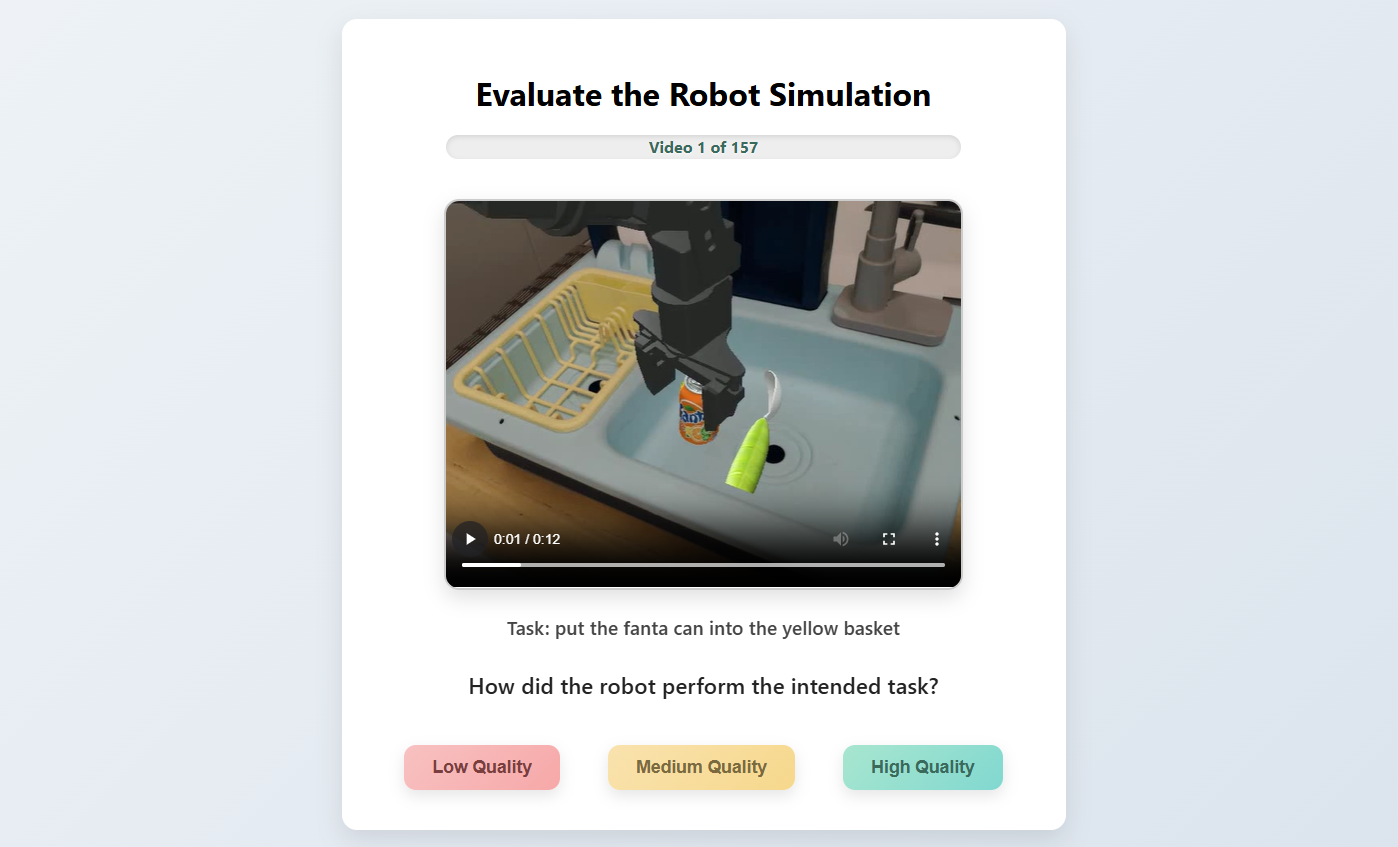}
    \caption{Screenshot of the web-based application for tagging}
    \label{fig:questionnaire}
\end{figure}

Two engineers with domain expertise were selected for labeling the success cases as (1) high-quality, (2) medium quality, and (3) low quality. The first engineer was a computer science engineer with a master's degree in robotics and autonomous systems, and had 2 years of post-master experience. The second engineer held two engineering degrees (electronics engineering degree and bachelor of science in industrial automation), as well as a master's degree in embedded systems, and had 13 years of post-master's studies. A total of 908 successful test cases were labeled. To reduce the fatigue of the labeling, multiple sessions across 15 days were conducted, with the maximum session involving 160 test cases to label. The agreement on labeling was assessed using Cohen's Kappa, yielding an agreement degree of 85\%, indicating almost perfect agreement. Disagreements were broken by a third labeler, also with domain expertise. To ease labeling, we developed a web-based approach that provided the domain experts with the video and instruction prompt for the tasks, in addition to the three options for tagging the quality level of the successful task. Figure~\ref{fig:questionnaire} shows an example of our web-based application for tagging. 

\begin{table*}[h!]
\centering
\caption{Quality level definition for each task}
\label{tab:Quality_levels}
\resizebox{\textwidth}{!}{%
\begin{tabular}{{clp{0.85\textwidth}}}
\toprule
\textbf{TASK} & \multicolumn{1}{c}{\textbf{QUALITY}} & \multicolumn{1}{c}{\textbf{CRITERIA}} \\ \cmidrule{1-3}
\multirow{6}{*}{\textbf{Pick up}} & \multirow{2}{*}{High} & The robot moves directly to the object, picks the target object up confidently, without hesitation or dropping. The object remains stable during the lift. \\ \cmidrule{2-3}
 & \multirow{2}{*}{Medium} & Some hesitation or one drop of the target object may occur. The object is picked up either in the first or at the second try, and may rotate or cause instability in the robot's movement. \\ \cmidrule{2-3}
 & \multirow{2}{*}{Low} & Significant delay, multiple drops or repeated failed attempts when lifting the object. The object may dropped, and not picked up again after lifting.  \\ \cmidrule{1-3}
\multirow{9}{*}{\textbf{Move near}} & \multirow{2}{*}{High} & Smooth and confident motion of the robot is seen, directly approaching the target object. Collisions or object drops are not produced. The object is grasped on the first attempt and maintains original orientation during and after task execution. \\ \cmidrule{2-3}
 & \multirow{3}{*}{Medium} & Slight pauses or inefficiencies may be observed in the robot motion. The robot is permitted to collide with surrounding objects maximum once or the target object being dropped maximum once. The grasping may require a maximum of two attempts. The orientation of the object is preserved during and after task execution. \\ \cmidrule{2-3}
 & \multirow{3}{*}{Low} & Multiple erroneous situations may occur, such as, repeated collisions, re-planning of the trajectory, or environmental disruption (e.g., multiple object collisions). More than two object drops may be needed before grasping the object. The orientation of the object may change during and after task execution.  \\ \cmidrule{1-3}
\multirow{7}{*}{\textbf{Put in}} & \multirow{2}{*}{High} & The object is accurately placed into the destination object on the first try. There is no hesitation or collision and minor adjustments are allowed. The are no unnecessary actions after placement.  \\ \cmidrule{2-3}
 & \multirow{2}{*}{Medium} & There may be minor corrections or the robot may fail on first try to grasp the target object but recovers quickly; for instance, the object can be re-grasped and correctly placed. Slight collisions with other objects or surroundings may occur.   \\ \cmidrule{2-3}
 & \multirow{2}{*}{Low} &  Multiple drops as well as failed attempts occur, or small pauses in execution where the robot stops to compute or replan the next movement. Abrupt motions or major collisions may occur and the object may not end up properly inside the target object. \\ \cmidrule{1-3}
\multirow{7}{*}{\textbf{Put on}} & \multirow{2}{*}{High} & The object is confidently and stably placed on the target in the first try. The motion is smooth and the orientation of the object is preserved. \\ \cmidrule{2-3}
 & \multirow{2}{*}{Medium} &  Minor corrections or slight collisions may occur. The object may slide when placing  or be partially placed, but finally adjusted. The orientation of the object is preserved. \\ \cmidrule{2-3}
 & \multirow{2}{*}{Low} &  The object can be dropped several times or repeatedly misplaced. Multiple corrections can be required or the object may end in an unstable or incorrect position on the target object. \\ \bottomrule
\end{tabular}%
}
\end{table*}

The definition of quality levels was defined before the execution, together with robotics domain experts. Labelers were instructed with clear and illustrative examples. We aimed to define these levels as objectively as possible for each of the four selected tasks. Table~\ref{tab:Quality_levels} provides our definition for each quality level for each task.

\subsection{Statistical tests}

In RQ2, we statistically measured the correlation between different successful task qualities and our metrics by using Spearman's rank correlation. A p-value below 0.05 was considered indicative of statistical significance. Positive correlation implied that our metrics could capture divergence in task quality. According to Cohen~\cite{cohen2013statistical} we can interpret the results as \textit{no correlation} if $\rho$ $<$ 0.10; \textit{weak correlation} if 0.10 $\leq$ $\rho$ $\leq$ 0.29; \textit{moderate correlation} if 0.29 $<$ $\rho$ $\leq$ 0.49; \textit{strong correlation} if $\rho$ $>$ 0.49.

\revision{Since each task execution produced a temporal sequence of metric values, we first aggregated these values into a single trajectory-level value. Specifically, for an execution of length $T$, we computed the mean metric value:}
\begin{equation}
\bar{m} = \frac{1}{T} \sum_{t=1}^{T} m_t
\end{equation}
\revision{\noindent where $m_t$ denotes the metric value at time step $t$.}

\revision{This aggregation produces one scalar value per execution, thereby reducing sensitivity to short-term fluctuations and irrelevant noise. The resulting value was then used to align the time-varying metric with the single human-annotated quality label (i.e., low, medium or high quality) for that trajectory. Spearman's rank correlation was subsequently computed across all executions between the aggregated metric values and the task quality levels.}

In RQ3, we measured the difference between the results achieved by each of the successful quality levels and the failing tasks. To do so, we first analyzed the data distribution using the Shapiro-Wilk test. Since the data was not normally distributed, we employed the \revision{Mann Whitney-U test}. A p-value below 0.05 was considered indicative of statistical significance between our metrics in each task quality and unsuccessful tasks. Additionally, we evaluated effect sizes through the Vargha and Delaney’s \^{A}$_{12}$ value. According to Romano et al.~\cite{romano2006exploring}, the effect size of the \^{A}$_{12}$ value can be categorized as \textit{negligible} if $d  < 0.147$, \textit{small} if $d <$ 0.33 , \textit{medium} if $d < 0.474$ and \textit{large} if $d \geq 0.474$ , where $d = 2|$\^{A}$_{12}$$ - 0.5|$.

For both, RQ2 and RQ3, our metrics provide a single value for each time step (as defined in Sections~\ref{sec:UncertaintyMetrics} and~\ref{sec:QualityMetrics}). To conduct the statistical tests, for each metric, we aggregated the values of each of time step and computed their average values over each run. 
This provides a single value per metric for each run, thereby enabling quantitative comparison across different experimental conditions.

\section{Analysis of the Results and Discussion}
\label{sec:results}

In this section we present a detailed analysis of the experimental results corresponding to our research questions. 

\subsection{RQ1 - Replication}
RQ1 replicates \revision{the experimental setup of} Wang et al. \cite{wang2025vlatest}, with  \revision{one key modification}: we manually annotated all successful tasks, with high-, medium-, or low-quality  \revision{labels. This allowed us to evaluate not only task success but also the actual quality of the task execution.} Furthermore,  \revision{after the annotation process} we identified some false negatives  \revision{raised by the symbolic oracle used in VLATest}, i.e., tasks labeled by the test oracle as \textit{``success''} that should not have been labeled as such (e.g., due to the object already being positioned inside the target object).

Table \ref{tab:RQ1Results}  \revision{presents} the overall results for RQ1. In terms of success rate  \revision{alone}, \pimodel{} and SpatialVLA generally outperformed OpenVLA,  \revision{whose} success rates ranged from 1.2\% to 13.8\%. SpatialVLA was the best model in the \textit{Pick up} and \textit{Move near} tasks, achieving 38.2 and 20.8\% of success rates, respectively. In contrast, \pimodel{} showed better performance in the \textit{Put In} and \textit{Put on} tasks with success rates of 14.4 and 18.6\%, respectively. For the first two tasks, the Google robot was used, whereas the latter employed the WidowX robot.  \revision{These results suggest that robot embodiment} may influence the overall performance of the task.

\begin{FindingBox}
When considering success rate alone, SpatialVLA was the best model in the first two tasks (involving the Google robot), whereas \pimodel{} performed best in the last two tasks (involving the WidowX robot).
\end{FindingBox}

\begin{table*}[h!]
\centering
\caption{RQ1 -- Number and ratio of successful task executions, along with their quality breakdown across different VLAs and tasks. False negatives refer to tasks labeled by the test oracle as \textit{``successful''}, but that should not have been labeled as such (e.g., when the target object was already in the destination position when the environment was set up).}
\label{tab:RQ1Results}

\begin{tabular}{llccccc}
\toprule
\textbf{Task} & \textbf{Model} & \textbf{Success} & \textbf{High-Quality} & \textbf{Medium-Quality} & \textbf{Low-Quality} & \textbf{False Neg.} \\
\midrule
\multirow{3}{*}{Pick up}   & OpenVLA     & 32 (6.4\%)  & 18 (56.2\%)  & 9 (28.1\%)   & 5 (15.7\%)   & 0 (0.0\%)  \\
          & \pimodel       & 161 (32.2\%)& 51 (31.7\%)  & 26 (16.1\%)  & 84 (52.2\%)  & 0 (0.0\%)  \\
          & SpatialVLA  & 191 (38.2\%)& 132 (69.1\%) & 31 (16.2\%)  & 28 (14.7\%)  & 0 (0.0\%)  \\
\midrule
\multirow{3}{*}{Move near} & OpenVLA     & 69 (13.8\%) & 7 (10.1\%)   & 21 (30.4\%)  & 41 (59.4\%)  & 0 (0.0\%)  \\
          & \pimodel       & 60 (12.0\%) & 16 (26.7\%)  & 9 (15.0\%)   & 35 (58.3\%)  & 0 (0.0\%)  \\
          & SpatialVLA  & 104 (20.8\%)& 55 (52.9\%)  & 17 (16.3\%)  & 32 (30.8\%)  & 2 (0.4\%)  \\
\midrule
\multirow{3}{*}{Put In}    & OpenVLA     & 9 (1.8\%)   & 5 (55.6\%)   & 3 (33.3\%)   & 1 (11.1\%)   & 10 (2.0\%) \\
          & \pimodel       & 72 (14.4\%) & 25 (34.7\%)  & 28 (38.9\%)  & 19 (26.4\%)  & 11 (2.2\%) \\
          & SpatialVLA  & 42 (8.4\%)  & 24 (57.1\%)  & 16 (38.1\%)  & 2 (4.8\%)    & 11 (2.2\%) \\
\midrule
\multirow{3}{*}{Put on}    & OpenVLA     & 6 (1.2\%)   & 2 (33.3\%)   & 2 (33.3\%)   & 2 (33.3\%)   & 13 (2.6\%) \\
          & \pimodel       & 93 (18.6\%) & 39 (41.9\%)  & 16 (17.2\%)  & 38 (40.9\%)  & 17 (3.4\%) \\
          & SpatialVLA  & 69 (13.8\%) & 30 (43.5\%)  & 19 (27.5\%)  & 20 (29.0\%)  & 15 (3.0\%) \\
\bottomrule
\end{tabular}
\end{table*}

However, when considering task quality,  \revision{the results were different. We found that SpatialVLA performs significantly better than \pimodel since the majority of the successful tasks from SpatialVLA were labeled by the annotators as ``\textit{high-quality}'', ranging from 43.5\% to 69.1\%, depending on the task type.} In contrast, for the \pimodel{} model, annotators classified many successful tasks as low-quality. This was especially concerning for the ``\textit{Pick up}'' and ``\textit{Move near}'' tasks, where more than half of the successful tasks were considered of low-quality. This suggests that the success rate metric alone \revision{is not enough to assess the task execution performance of VLA-enabled robots, highlighting a critical limitation of the symbolic oracle used in VLATest~\cite{wang2025vlatest}.} For instance, \pimodel{}  successfully completed the ``\textit{Pick up}'' task 161 times, closely approaching SpatialVLA, which achieved 191 successful executions. However, according to the annotators' classification, \pimodel{} only achieved 51 high-quality executions, whereas SpatialVLA more than doubled this performance with 132 high-quality successful executions. Even in the cases where \pimodel{} was better, the quality of its executions was not very high. For instance, in the ``\textit{Put in}'' task, \pimodel{} obtained a total of 72 successful executions compared to SpatialVLA's 42. However, annotators judged \pimodel{} to have only one more high-quality execution than SpatialVLA (25 vs 24). 

Lastly, the quality of OpenVLA varied by task. In the ``\textit{Pick up}'' task, most successful executions were labeled as high-quality. Conversely, for the ``\textit{Move near}'' task, most were of low-quality. The success rates for OpenVLA in the last two tasks were too low to draw any reliable conclusions.

\begin{FindingBox}
\revision{Success rate alone is not a sufficient metric for testing VLA-enabled robots. Similar success rates can still exhibit large differences in the task quality, exposing the weaknesses of binary success-based testing approaches.} SpatialVLA was found to be the VLA with the highest overall quality of task execution.
\end{FindingBox}

\subsection{RQ2 - Correlation}

Table \ref{tab:Spearman2} shows the overall correlation between the quality levels assigned by the annotators and the uncertainty and quality metrics proposed in Sections \ref{sec:UncertaintyMetrics} and \ref{sec:QualityMetrics}, respectively. \revision{For both Action-based and Tool Center Point-based metrics, we distinguish three variants: 1) the full metric, 2) the metric computed using only the positional component of the action or position, and 3) the metric computed using only the orientation component. This distinction is motivated by the fact that, in VLA-enabled robots, positional and orientation errors can have fundamentally different impacts on task success and perceived quality i.e., the same position can be reached using different orientations. Separating these components allows for a more fine-grained analysis of model behavior, helping to identify whether uncertainty or performance degradation arises primarily from spatial inaccuracies or from misalignment in orientation.} When measuring the correlation between the quality classification performed by the annotators and the uncertainty and quality metrics proposed in this paper, results differed based on the task type and analyzed VLA model.

\begin{table*}[ht]
\caption{\revision{Spearman rank correlation coefficients between the metrics and human evaluation results for each task type. Higher correlation ($\rho$) values indicate stronger monotonic relationships; that is, lower metric values correspond to higher human-evaluated task quality. Cells with statistically significant results (i.e., $p < 0.05$) are highlighted in blue. We interpret the correlation as follows~\cite{cohen2013statistical}: \colorbox{cyan!7}{\textit{no correlation}} if $\rho$ $<$ 0.10; \colorbox{cyan!20}{\textit{weak correlation}} if 0.10 $\leq$ $\rho$ $\leq$ 0.29; \colorbox{cyan!40}{\textit{moderate correlation}} if 0.29 $<$ $\rho$ $\leq$ 0.49; \colorbox{cyan!85}{\textit{strong correlation}} if $\rho$ $>$ 0.49.}}
\label{tab:Spearman2}
\resizebox{\textwidth}{!}{%
\begin{tabular}{llrrrrrrrrrrrrr}
\toprule
\multicolumn{1}{l}{} & & \multicolumn{3}{c}{\textbf{Pick up}} & \multicolumn{3}{c}{\textbf{Move near}} & \multicolumn{3}{c}{\textbf{Put in}} & \multicolumn{3}{c}{\textbf{Put on}} \\ \cmidrule{3-14} 
\multicolumn{1}{l}{} &  & \multicolumn{1}{c}{OpenVLA} & \multicolumn{1}{c}{\pimodel} & \multicolumn{1}{c}{SpatialVLA} & \multicolumn{1}{c}{OpenVLA} & \multicolumn{1}{c}{\pimodel} & \multicolumn{1}{c}{SpatialVLA} & \multicolumn{1}{c}{OpenVLA} & \multicolumn{1}{c}{\pimodel} & \multicolumn{1}{c}{SpatialVLA} & \multicolumn{1}{c}{OpenVLA} & \multicolumn{1}{c}{\pimodel} & \multicolumn{1}{c}{SpatialVLA} \\ \cmidrule{1-14} 

 \multirow{16}{*}{\textbf{Uncertainty Metrics}} & TB-TP  & \cellcolor{cyan!40}{0.435} & \cellcolor{cyan!85}{0.602} & \cellcolor{cyan!85}{0.551} & -0.094 & \cellcolor{cyan!40}{0.324} & \cellcolor{cyan!40}{0.421} & -0.620 & 0.007 & \cellcolor{cyan!85}{0.569} & 0.717 & \cellcolor{cyan!7}{-0.657} & \cellcolor{cyan!40}{0.330} \\
 & TB-PCS  & \cellcolor{cyan!40}{0.435} & \cellcolor{cyan!40}{0.376} & \cellcolor{cyan!85}{0.530} & -0.094 & \cellcolor{cyan!40}{0.295} & \cellcolor{cyan!40}{0.396} & -0.507 & -0.047 & \cellcolor{cyan!40}{0.472} & 0.717 & \cellcolor{cyan!7}{-0.572} & \cellcolor{cyan!40}{0.338} \\
 & TB-D  & \cellcolor{cyan!40}{0.457} & - & \cellcolor{cyan!85}{0.575} & -0.092 & - & \cellcolor{cyan!40}{0.454} & -0.620 & - & \cellcolor{cyan!85}{0.549} & 0.717 & - & \cellcolor{cyan!40}{0.331} \\
 & TB-E  & \cellcolor{cyan!40}{0.443} & \cellcolor{cyan!20}{0.173} & \cellcolor{cyan!85}{0.597} & -0.079 & \cellcolor{cyan!40}{0.360} & \cellcolor{cyan!40}{0.482} & -0.620 & -0.084 & \cellcolor{cyan!85}{0.547} & \cellcolor{cyan!85}{0.837} & \cellcolor{cyan!40}{0.373} & \cellcolor{cyan!40}{0.318} \\
 & A-PI$_{all}$ & \cellcolor{cyan!40}{0.395} & \cellcolor{cyan!85}{0.829} & \cellcolor{cyan!85}{0.624} & 0.147 & 0.139 & \cellcolor{cyan!40}{0.333} & -0.394 & 0.080 & \cellcolor{cyan!40}{0.346} & 0.717 & \cellcolor{cyan!85}{0.762} & \cellcolor{cyan!40}{0.415} \\
 & A-PI$_{o}$  & \cellcolor{cyan!85}{0.560} & \cellcolor{cyan!85}{0.739} & \cellcolor{cyan!85}{0.576} & 0.025 & \cellcolor{cyan!40}{0.427} & \cellcolor{cyan!40}{0.410} & -0.620 & \cellcolor{cyan!20}{0.247} & \cellcolor{cyan!40}{0.380} & 0.598 & \cellcolor{cyan!85}{0.598} & \cellcolor{cyan!20}{0.255} \\
 & A-PI$_{p}$  & \cellcolor{cyan!85}{0.501} & \cellcolor{cyan!85}{0.728} & \cellcolor{cyan!85}{0.552} & 0.028 & \cellcolor{cyan!40}{0.312} & \cellcolor{cyan!40}{0.376} & -0.169 & \cellcolor{cyan!40}{0.353} & \cellcolor{cyan!85}{0.502} & \cellcolor{cyan!85}{0.956} & \cellcolor{cyan!85}{0.546} & 0.186 \\
   & A-VI$_{all}$  & \cellcolor{cyan!40}{0.412} & \cellcolor{cyan!85}{0.833} & \cellcolor{cyan!85}{0.608} & 0.181 & 0.131 & \cellcolor{cyan!20}{0.285} & -0.394 & 0.077 & \cellcolor{cyan!40}{0.348} & \cellcolor{cyan!85}{0.837} & \cellcolor{cyan!85}{0.768} & \cellcolor{cyan!40}{0.411} \\
   & A-VI$_{o}$  & \cellcolor{cyan!85}{0.565} & \cellcolor{cyan!85}{0.716} & \cellcolor{cyan!85}{0.598} & 0.035 & \cellcolor{cyan!40}{0.443} & \cellcolor{cyan!40}{0.434} & -0.507 & 0.185 & \cellcolor{cyan!40}{0.375} & 0.717 & \cellcolor{cyan!85}{0.610} & \cellcolor{cyan!20}{0.239} \\
 & A-VI$_{p}$  & \cellcolor{cyan!85}{0.529} & \cellcolor{cyan!85}{0.673} & \cellcolor{cyan!85}{0.540} & 0.069 & \cellcolor{cyan!40}{0.361} & \cellcolor{cyan!40}{0.388} & -0.056 & \cellcolor{cyan!40}{0.291} & \cellcolor{cyan!85}{0.500} & \cellcolor{cyan!85}{0.956} & \cellcolor{cyan!85}{0.533} & 0.157 \\
 & A-AI$_{all}$  & \cellcolor{cyan!40}{0.423} & \cellcolor{cyan!85}{0.828} & \cellcolor{cyan!85}{0.595} & 0.176 & 0.114 & \cellcolor{cyan!20}{0.263} & -0.394 & 0.09 & \cellcolor{cyan!40}{0.352} & \cellcolor{cyan!85}{0.837} & \cellcolor{cyan!85}{0.768} & \cellcolor{cyan!40}{0.408} \\
 & A-AI$_{o}$  & \cellcolor{cyan!85}{0.579} & \cellcolor{cyan!85}{0.697} & \cellcolor{cyan!85}{0.600} & 0.024 & \cellcolor{cyan!40}{0.450} & \cellcolor{cyan!40}{0.438} & -0.507 & 0.172 & \cellcolor{cyan!40}{0.390} & 0.717 & \cellcolor{cyan!85}{0.605} & \cellcolor{cyan!20}{0.239} \\
 & A-AI$_{p}$  & \cellcolor{cyan!85}{0.543} & \cellcolor{cyan!85}{0.639} & \cellcolor{cyan!85}{0.561} & 0.056 & \cellcolor{cyan!40}{0.361} & \cellcolor{cyan!40}{0.402} & -0.056 & \cellcolor{cyan!20}{0.270} & \cellcolor{cyan!40}{0.439} & \cellcolor{cyan!85}{0.956} & \cellcolor{cyan!85}{0.526} & 0.137 \\
 & EV  & - & \cellcolor{cyan!40}{0.357} & - &  & 0.203 & - &  & -0.077 & -0.248 & - & \cellcolor{cyan!85}{0.680} & -0.173 \\
 \cmidrule{1-14}
 \multirow{11}{*}{\textbf{Quality Metrics}}   & TCP-PI$_{all}$  & \cellcolor{cyan!85}{0.523} & \cellcolor{cyan!85}{0.576} & -0.019 & 0.124 & \cellcolor{cyan!40}{0.407} & -0.015 & 0.169 & \cellcolor{cyan!40}{0.294} & 0.284 & \cellcolor{cyan!85}{0.837} & \cellcolor{cyan!85}{0.549} & \cellcolor{cyan!20}{0.289} \\
   & TCP-PI$_{o}$  & \cellcolor{cyan!85}{0.599} & \cellcolor{cyan!85}{0.507} & \cellcolor{cyan!7}{-0.281} & 0.153 & \cellcolor{cyan!40}{0.456} & -0.051 & 0.394 & \cellcolor{cyan!20}{0.264} & 0.228 & 0.598 & \cellcolor{cyan!85}{0.544} & \cellcolor{cyan!40}{0.292} \\
 & TCP-PI$_{p}$  & \cellcolor{cyan!40}{0.421} & \cellcolor{cyan!85}{0.586} & \cellcolor{cyan!40}{0.315} & 0.044 & \cellcolor{cyan!40}{0.300} & 0.115 & -0.056 & \cellcolor{cyan!20}{0.270} & \cellcolor{cyan!40}{0.368} & \cellcolor{cyan!85}{0.837} & \cellcolor{cyan!40}{0.409} & \cellcolor{cyan!20}{0.239} \\
 & TCP-VI$_{all}$ & \cellcolor{cyan!85}{0.651} & \cellcolor{cyan!85}{0.699} & \cellcolor{cyan!85}{0.561} & 0.066 & \cellcolor{cyan!40}{0.383} & \cellcolor{cyan!40}{0.423} & 0.394 & \cellcolor{cyan!20}{0.261} & \cellcolor{cyan!85}{0.500} & \cellcolor{cyan!85}{0.837} & \cellcolor{cyan!85}{0.597} & \cellcolor{cyan!40}{0.304} \\
 & TCP-VI$_{o}$  & \cellcolor{cyan!85}{0.693} & \cellcolor{cyan!85}{0.660} & \cellcolor{cyan!85}{0.541} & 0.088 & \cellcolor{cyan!40}{0.411} & \cellcolor{cyan!40}{0.440} & 0.394 & 0.231 & \cellcolor{cyan!40}{0.452} & 0.717 & \cellcolor{cyan!85}{0.595} & \cellcolor{cyan!40}{0.311} \\
 & TCP-VI$_{p}$  & \cellcolor{cyan!85}{0.536} & \cellcolor{cyan!85}{0.697} & \cellcolor{cyan!85}{0.547} & 0.067 & \cellcolor{cyan!40}{0.301} & \cellcolor{cyan!40}{0.344} & 0.394 & 0.211 & \cellcolor{cyan!85}{0.637} & \cellcolor{cyan!85}{0.837} & \cellcolor{cyan!40}{0.467} & 0.236 \\
 
 & TCP-AI$_{all}$  & \cellcolor{cyan!85}{0.657} & \cellcolor{cyan!85}{0.684} & \cellcolor{cyan!85}{0.546} & 0.03 & \cellcolor{cyan!40}{0.389} & \cellcolor{cyan!40}{0.461} & 0.394 & \cellcolor{cyan!20}{0.246} & \cellcolor{cyan!85}{0.504} & \cellcolor{cyan!85}{0.837} & \cellcolor{cyan!85}{0.617} & \cellcolor{cyan!20}{0.246} \\
  & TCP-AI$_{o}$  & \cellcolor{cyan!85}{0.682} & \cellcolor{cyan!85}{0.653} & \cellcolor{cyan!85}{0.523} & 0.046 & \cellcolor{cyan!40}{0.424} & \cellcolor{cyan!40}{0.464} & 0.282 & 0.214 & \cellcolor{cyan!40}{0.456} & \cellcolor{cyan!85}{0.837} & \cellcolor{cyan!85}{0.623} & \cellcolor{cyan!20}{0.260} \\
 & TCP-AI$_{p}$  & \cellcolor{cyan!85}{0.592} & \cellcolor{cyan!85}{0.664} & \cellcolor{cyan!85}{0.539} & 0.047 & \cellcolor{cyan!40}{0.306} & \cellcolor{cyan!40}{0.406} & 0.394 & 0.190 & \cellcolor{cyan!85}{0.613} & 0.717 & \cellcolor{cyan!85}{0.514} & 0.179 \\
 & TI  & \cellcolor{cyan!85}{0.531} & \cellcolor{cyan!85}{0.665} & \cellcolor{cyan!85}{0.527} & 0.064 & \cellcolor{cyan!40}{0.355} & \cellcolor{cyan!40}{0.379} & 0.394 & 0.226 & \cellcolor{cyan!85}{0.575} & 0.717 & \cellcolor{cyan!40}{0.473} & 0.205 \\
 & OT  & \cellcolor{cyan!40}{0.414} & \cellcolor{cyan!85}{0.540} & \cellcolor{cyan!40}{0.327} & \cellcolor{cyan!40}{0.386} & \cellcolor{cyan!40}{0.327} & \cellcolor{cyan!7}{-0.390} & 0.169 & -0.06 & 0.085 & -0.478 & \cellcolor{cyan!7}{-0.217} & \cellcolor{cyan!7}{-0.252} \\

 \bottomrule

\end{tabular}%
}
\end{table*}

For the uncertainty metrics, nearly all showed a statistical significant positive correlation for SpatialVLA. Moreover, this correlation was high or moderate in most cases and tasks based on the levels from~\cite{cohen2013statistical}. The A-PI, A-VI, and A-AI uncertainty metrics showed strong statistically significant correlations for the \textit{``Pick up''} and \textit{``Put on''} tasks in the case of the \pimodel. In contrast, for the \pimodel, no metrics showed positive correlation for the ``\textit{Put in}'' task, while only three of them (TB-TP, TB-PCS and TB-E) showed moderate correlation with statistical significance for the \textit{``Move near''} task. Lastly, for OpenVLA, all uncertainty metrics showed statistically significant moderate correlation for the \textit{``Pick up''} task, but no correlation for the \textit{``Move near''} task. The sample size was too small for this model to enable any meaningful statistical assessment for the \textit{``Put in''} and \textit{``Put on''} tasks. 

\revision{When comparing the positional ($p$) and orientation ($o$) components of the action-based metrics, the ones measured only with the orientation component frequently achieved equal or higher correlations than their positional counterparts. This trend was particularly evident for the \textit{``Pick up''} task, where A-PI$_o$, A-VI$_o$, and A-AI$_o$ generally outperformed A-PI$_p$, A-VI$_p$, and A-AI$_p$ across the three models. A similar patter can be observed for the \textit{``Move near''} task in the case of \pimodel and SpatialVLA. These results suggest that uncertainty associated with the predicted end-effector orientation is often more closely aligned with human perception of task quality than uncertainty in position alone. Nevertheless, neither component consistently dominated across all tasks, indicating that positional and orientation information are complementary for uncertainty estimation.}

Overall, the uncertainty metrics A-VI and A-AI seemed to be the most appropriate metrics, showing moderate to strong correlation in 8 out of 12 task-model combinations. \revision{Furthermore, the strong performance of both their positional and orientation components suggests that these metrics are able to capture quality-relevant aspects of the robot motion regardless of whether errors arise primarily from spatial inaccuracies or orientation misalignment.} In contrast, EV showed statistically  significant only for the \textit{``Pick up''} task with the \pimodel{} VLA model. In 6 out of 12 tasks, for the EV metrics, correlation could not be measured because the different model instances provided the exact same inference values in all cases. For the remaining five cases, there was no statistically significant correlation. Therefore, we do not recommend using this metric for measuring uncertainty.

\begin{FindingBox}
The performance of uncertainty metrics depends on the specific task and VLA model used. Overall, A-VI and A-AI were reliable metrics, showing moderate to high correlation in many tasks. \revision{In addition, the correlation of orientation-based variants highlighted the importance of considering orientation uncertainty when evaluating VLA behavior.} However, we recommend that practitioners perform an in-depth analysis of the metrics in the context of their specific models.
\end{FindingBox}

In the case of the quality metrics, results also differed depending on the VLA model and task type. Similar to the uncertainty metrics, for the \textit{``Pick up''} task, all available metrics showed statistically significant correlation. Specifically, TCP-VI, TCP-AI, and TI showed strong correlation for all three models; whereas, TCP-PI and OT showed strong correlation for \pimodel{} and moderate correlation for OpenVLA and SpatialVLA models. For the \textit{``Move near''} task, all metrics showed moderate correlation for the \pimodel{} models, whereas only the TCP-VI, TCP-AI, and TI metrics showed moderate correlation for SpatialVLA. For OpenVLA, only OT showed moderate correlation, whereas the rest did not have any correlation with statistical significance. For the last two tasks, OpenVLA obtained very few successful executions (9 and 6, respectively). Therefore, we believe that the sample size is too small to make any statistically sound claims for these tasks and model. All metrics except the OT showed statistically significant positive correlation for SpatialVLA for the \textit{``Put in''} task, three of which (TCP-VI, TCP-AI, and TI) had strong correlation. Meanwhile, only TCP-PI had a statistically significant positive correlation for \pimodel{} and this third task. Conversely, for the \textit{``Put on''} task, all metrics except for the OT had moderate to strong correlation for \pimodel. In contrast, only TCP-VI showed positive correlation with statistical significance in this same task for SpatialVLA. 

\revision{A comparison of the metrics measured only on the orientation and positional components of the TCP-based metrics revealed a more task-dependent behavior that the observed for uncertainty estimation. For the \textit{``Pick up''} task, orientation-based metric frequently achieved the highest correlations, indicating that smooth and consistent end-effector orientation plays an important role in perceived grasp quality. In contrast, for the \textit{``Put in''} task, positional-based metrics often outperformed orientation ones, particularly for SpatialVLA, suggesting that accurately reaching the desired location is more important than the orientation in these kind of tasks. The combined variants generally remained among the best-performing metrics, supporting the view that both capure complementary aspects of execution quality.}

Overall, TCP-VI seems to be the most consistent metric in achieving statistically significant positive correlation between successful tasks of varying quality. Smooth and uninterrupted motion reflects precise and confident control, whereas velocity spikes reveal hesitation or corrective errors that degrade performance and task quality. This means that instability in the velocity of the end-effector point of the robot seems to be an appropriate metric to measure task quality on VLA models. \revision{Interestingly, TCP-VI remained robust regardless of whether it was computed using positional, orientation, or combined information, further supporting its suitability as a general-purpose quality metric.}

\begin{FindingBox}
Similar to the uncertainty metrics, the performance of quality metrics depends on the specific task and VLA model. Overall, TCP-VI seems to be the most consistent metric across models and tasks and is therefore recommended for use. 
\end{FindingBox}

\begin{table}[ht]
\centering
\caption{\revision{Spearman's rank correlation ($\rho$) coefficients between the metrics and human evaluation results for the \textit{Move near} task when removing the Low-quality tasks due to bad orientation of the object }}
\label{tab:moveNearCorr}
\resizebox{0.475\textwidth}{!}{
\begin{tabular}{llrrr}
\toprule
& & \textbf{OpenVLA} & \pimodel & \textbf{SpatialVLA} \\
\cmidrule{1-5}
\multirow{16}{*}{\textbf{Uncertainty Metrics}} & TB-TP & 0.069 & \colorbox{cyan!40}{0.331} & \colorbox{cyan!40}{0.392} \\
& TB-PCS & 0.079 & 0.262 & \colorbox{cyan!40}{0.365} \\
& TB-D & 0.071 & - & \colorbox{cyan!40}{0.423} \\
& TB-E & 0.067 & \colorbox{cyan!40}{0.447} & \colorbox{cyan!40}{0.456} \\
& A-PI$_{all}$ & 0.245 & 0.246 & \colorbox{cyan!40}{0.317} \\
& A-PI$_{o}$ & 0.069 & \colorbox{cyan!85}{0.512} & \colorbox{cyan!40}{0.436} \\
& A-PI$_{p}$ & 0.116 & \colorbox{cyan!40}{0.429} & \colorbox{cyan!40}{0.391} \\
& A-VI$_{all}$ & 0.281 & 0.236 & \colorbox{cyan!20}{0.268} \\
& A-VI$_{o}$ & 0.077 & \colorbox{cyan!85}{0.54} & \colorbox{cyan!40}{0.44} \\
& A-VI$_{p}$ & 0.174 & \colorbox{cyan!40}{0.482} & \colorbox{cyan!40}{0.403} \\
& A-AI$_{all}$ & 0.273 & 0.211 & \colorbox{cyan!20}{0.249} \\
& A-AI$_{o}$ & 0.067 & \colorbox{cyan!85}{0.543} & \colorbox{cyan!40}{0.434} \\
& A-AI$_{p}$ & 0.141 & \colorbox{cyan!40}{0.47} & \colorbox{cyan!40}{0.412} \\
& EV & - & \colorbox{cyan!40}{0.306} & - \\ \cmidrule{1-5}
\multirow{11}{*}{\textbf{Quality Metrics}}& TCP-PI$_{all}$ & 0.216 & \colorbox{cyan!85}{0.518} & 0.068 \\
& TCP-PI$_{o}$ & 0.275 & \colorbox{cyan!85}{0.539} & 0.026 \\
& TCP-PI$_{p}$ & 0.083 & \colorbox{cyan!40}{0.413} & 0.188 \\
& TCP-VI$_{all}$ & 0.261 & \colorbox{cyan!40}{0.481} & \colorbox{cyan!40}{0.444} \\
& TCP-VI$_{o}$ & \colorbox{cyan!20}{0.289} & \colorbox{cyan!40}{0.488} & \colorbox{cyan!40}{0.46} \\
& TCP-VI$_{p}$ & 0.240 & \colorbox{cyan!40}{0.411} & \colorbox{cyan!40}{0.376} \\
& TCP-AI$_{all}$ & 0.233 & \colorbox{cyan!40}{0.483} & \colorbox{cyan!40}{0.468} \\
& TCP-AI$_{o}$ & 0.243 & \colorbox{cyan!85}{0.502} & \colorbox{cyan!40}{0.472} \\
& TCP-AI$_{p}$ & 0.262 & \colorbox{cyan!40}{0.427} & \colorbox{cyan!40}{0.427} \\
& TI & 0.243 & \colorbox{cyan!40}{0.476} & \colorbox{cyan!40}{0.406} \\
& OT & \colorbox{cyan!40}{0.398} & \colorbox{cyan!40}{0.315} & \colorbox{cyan!7}{-0.355} \\ \bottomrule
\end{tabular}}
\end{table}

\begin{figure}[h!]
    \centering
    
    \begin{subfigure}[b]{0.45\textwidth}
        \centering
        \includegraphics[width=\linewidth]{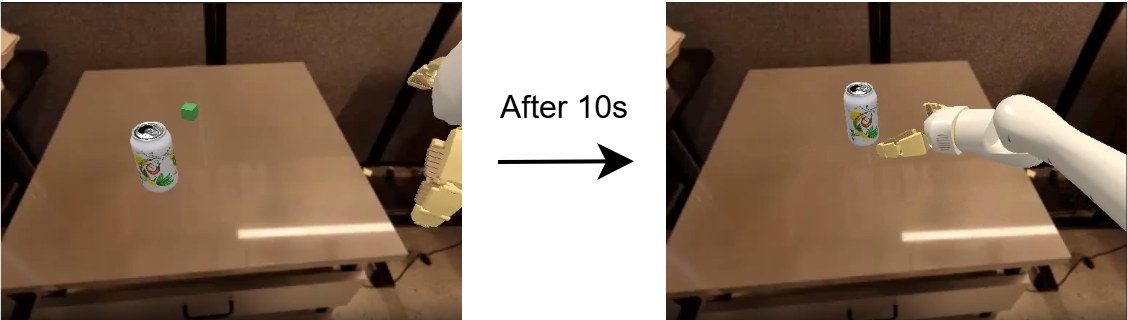} 
        \caption{Correct orientation of the object}
        \label{fig:good}
    \end{subfigure}
    \vspace{0.3em}
    \begin{subfigure}[b]{0.45\textwidth}
        \centering
        \includegraphics[width=\linewidth]{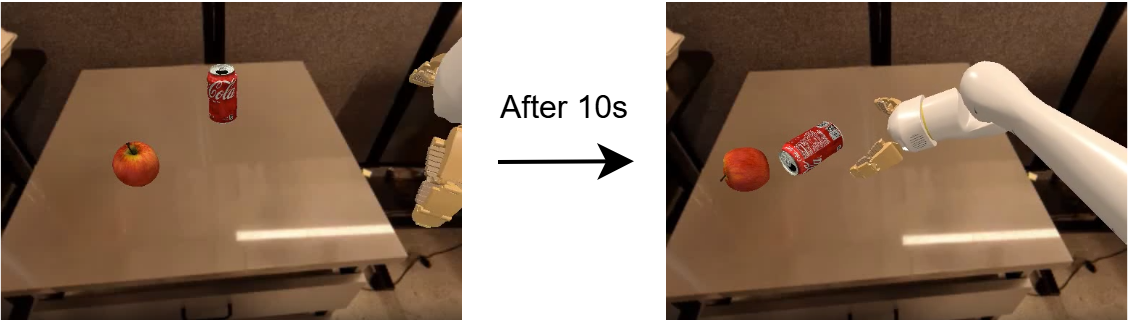} 
        \caption{Incorrect orientation of the object}
        \label{fig:bad}
    \end{subfigure}
    \caption{Comparison between correct and incorrect orientations when performing the \textit{Move near} task}
    \label{fig:orientation_comparison}
\end{figure}

Interestingly, for the \textit{``Move near''} task, some successful executions were labeled as ``low-quality'' if the final object orientation was not the correct, i.e., not the same or similar orientation to its original pose. Figure~\ref{fig:orientation_comparison} shows two examples of what we consider correct and incorrect orientation. After removing these cases from the data and recomputing the correlations, we found that both for \pimodel{} and SpatialVLA, most metrics showed strong correlation, as shown in Table~\ref{tab:moveNearCorr}.  
This may suggest that our metrics, both uncertainty and quality, are not able to adequately capture the correctness of object orientation. This pattern was not observed in the other tasks because the orientation was not considered critical for those scenarios. However, we considered it critical for the \textit{``Move near''} task, as it may cause problems, e.g., if the targeted object is an open water of bottle, it could spill. Since the simulators usually have access to both the initial and final positions, as well as orientation of the objects, a promising direction for improving our metrics would be to identify tasks where the orientation is important and, subsequently, integrate orientation in our metrics. This remains an open challenge that will be targeted in the future.

\begin{FindingBox}
Our metrics do not adequately capture the final orientation of the objects for the \textit{``Move near''} task, which we consider an important aspect in this specific task. Future research avenues could explore incorporating object orientation into our metrics to improve the performance. 
\end{FindingBox}

\subsection{RQ3 - Discrimination}

In this third RQ we studied the extent to which the \revision{uncertainty and quality} metrics can distinguish successful and failing task executions. Tables~\ref{tab:RQ3_OpenVLA}, \ref{tab:RQ3_pi0}, and \ref{tab:RQ3_spatialVLA} present statistical comparisons of the uncertainty and quality metrics between successful tasks of different quality levels and failing tasks. As observed in the aforementioned tables, the differences varied depending on the model, task, and classification of task success. 

\begin{table*}[h!]
\centering
\caption{\textcolor{black}{Comparison of $\hat{A}_{12}$ effect sizes for OpenVLA between successful and failing tasks across metrics, tasks, and quality levels. Each cell presents the Vargha–Delaney $\hat{A}_{12}$ effect size comparing a quality group (High, Medium, or Low) against failing results, for a specific task and metric. Cells are color-coded only if the comparison is statistically significant (i.e., $p < 0.05$) based on the Mann–Whitney U test. Cyan cells indicate that the quality group performed better than the failing group (i.e., $\hat{A}_{12}$ $< 0.5$), while red cells indicate the opposite (i.e., A12 $> 0.5$). Color intensity increases with effect size magnitude: light \colorbox{cyan!7}{cyan} and \colorbox{red!7}{red} for negligible effects ($d < 0.147$), moderate \colorbox{cyan!20}{cyan} and \colorbox{red!30}{red} for small ($d< 0.33$), darker \colorbox{cyan!40}{cyan} and \colorbox{red!50}{red} for medium ($d< 0.474$), and darkest \colorbox{cyan!85}{cyan} and \colorbox{red!70}{red} for large effects ($d\geq 0.474$), where $d = 2|\hat{A}_{12}- 0.5|$. Cells containing a ``-'' mean that the number of samples is insufficient to perform the statistical tests.
}}
\label{tab:RQ3_OpenVLA}
\resizebox{\textwidth}{!}{
\begin{tabular}{llcccccccccccc}
\toprule
& & \multicolumn{3}{c}{\textbf{Pick up}}& \multicolumn{3}{c}{\textbf{Move Near}}& \multicolumn{3}{c}{\textbf{Put in}}& \multicolumn{3}{c}{\textbf{Put on}}\\
\cmidrule{2-14}
& & High & Medium & Low & High & Medium & Low & High & Medium & Low & High & Medium & Low \\
\cmidrule{1-14}
\multirow{14}{*}{\textbf{Uncertainty Metrics}}& TB-TP & 0.53 & 0.62 & 0.72 & 0.62 & 0.62 & 0.58 & \cellcolor{cyan!85} 0.19 & \cellcolor{cyan!85} 0.05 & - & - & - & - \\
& TB-PCS & 0.53 & 0.62 & 0.73 & 0.62 & 0.62 & 0.59 & \cellcolor{cyan!85} 0.19 & \cellcolor{cyan!85} 0.05 & - & - & - & - \\
& TB-D & 0.53 & 0.62 & 0.72 & 0.61 & 0.62 & 0.58 & \cellcolor{cyan!85} 0.20 & \cellcolor{cyan!85} 0.05 & - & - & - & - \\
& TB-E & 0.52 & 0.62 & 0.71 & 0.61 & 0.62 & 0.58 & \cellcolor{cyan!85} 0.19 & \cellcolor{cyan!85} 0.05 & - & - & - & - \\
& A-PI$_{all}$ & 0.40 & 0.57 & 0.66 & \cellcolor{red!60} 0.72 & \cellcolor{red!60} 0.69 & \cellcolor{red!60} 0.72 & 0.48 & 0.43 & - & - & - & - \\
& A-PI$_{o}$ & 0.44 & 0.61 & 0.68 & 0.62 & 0.57 & \cellcolor{red!30} 0.60 & 0.46 & 0.19 & - & - & - & - \\
& A-PI$_{p}$ & 0.48 & 0.64 & 0.71 & 0.62 & \cellcolor{red!30} 0.64 & \cellcolor{red!30} 0.62 & 0.66 & 0.56 & - & - & - & - \\

& A-VI$_{all}$ & 0.40 & 0.57 & 0.65 & \cellcolor{red!60} 0.73 & \cellcolor{red!60} 0.69 & \cellcolor{red!60} 0.73 & 0.48 & 0.43 & - & - & - & - \\
& A-VI$_{o}$ & 0.43 & 0.61 & 0.72 & 0.61 & 0.56 & \cellcolor{red!30} 0.59 & 0.45 & 0.18 & - & - & - & - \\
& A-VI$_{p}$ & 0.45 & 0.62 & 0.74 & 0.59 & 0.62 & \cellcolor{red!30} 0.62 & 0.66 & 0.59 & - & - & - & - \\
& A-AI$_{all}$ & 0.40 & 0.57 & 0.65 & \cellcolor{red!60} 0.73 & \cellcolor{red!60} 0.69 & \cellcolor{red!60} 0.73 & 0.49 & 0.45 & - & - & - & - \\
& A-AI$_{o}$ & 0.42 & 0.62 & 0.72 & 0.62 & 0.56 & 0.59 & 0.45 & 0.20 & - & - & - & - \\
& A-AI$_{p}$ & 0.44 & 0.62 & 0.75 & 0.59 & 0.62 & \cellcolor{red!30} 0.61 & 0.66 & 0.60 & - & - & - & - \\
& EV & 0.50 & 0.50 & 0.50 & 0.50 & 0.50 & 0.50 & 0.50 & 0.50 & - & - & - & - \\
\cmidrule{1-14}
\multirow{11}{*}{\textbf{Quality Metrics}} & TCP-PI$_{all}$ & 0.56 & \cellcolor{red!60} 0.74 & 0.73 & 0.53 & 0.61 & \cellcolor{red!30} 0.63 & \cellcolor{cyan!85} 0.23 & 0.30 & - & - & - & - \\
& TCP-PI$_{o}$ & 0.52 & 0.68 & 0.72 & 0.51 & 0.61 & \cellcolor{red!30} 0.64 & 0.25 & 0.33 & - & - & - & - \\
& TCP-PI$_{p}$ & 0.62 & \cellcolor{red!85} 0.81 & 0.75 & 0.60 & \cellcolor{red!30} 0.63 & \cellcolor{red!30} 0.63 & 0.54 & 0.53 & - & - & - & - \\
& TCP-VI$_{all}$ & 0.46 & 0.66 & 0.73 & 0.56 & \cellcolor{red!30} 0.64 & \cellcolor{red!30} 0.63 & 0.27 & 0.52 & - & - & - & - \\
& TCP-VI$_{o}$ & 0.44 & 0.65 & 0.73 & 0.56 & \cellcolor{red!30} 0.65 & \cellcolor{red!30} 0.64 & 0.27 & 0.54 & - & - & - & - \\
& TCP-VI$_{p}$ & 0.50 & \cellcolor{red!60} 0.69 & 0.70 & 0.60 & \cellcolor{red!30} 0.66 & \cellcolor{red!30} 0.63 & 0.43 & 0.68 & - & - & - & - \\
& TCP-AI$_{all}$ & 0.45 & 0.65 & 0.74 & 0.58 & \cellcolor{red!30} 0.66 & \cellcolor{red!30} 0.64 & 0.27 & 0.54 & - & - & - & - \\
& TCP-AI$_{o}$ & 0.44 & 0.63 & 0.74 & 0.57 & \cellcolor{red!60} 0.67 & \cellcolor{red!30} 0.64 & 0.26 & 0.56 & - & - & - & - \\
& TCP-AI$_{p}$ & 0.48 & 0.69 & 0.72 & 0.62 & \cellcolor{red!60} 0.67 & \cellcolor{red!30} 0.64 & 0.40 & 0.66 & - & - & - & - \\
& TI & 0.49 & 0.66 & 0.71 & 0.58 & \cellcolor{red!30} 0.64 & \cellcolor{red!30} 0.61 & 0.40 & 0.67 & - & - & - & - \\
& OT & \cellcolor{cyan!85} 0.06 & \cellcolor{cyan!85} 0.08 & 0.33 & \cellcolor{cyan!85} 0.01 & \cellcolor{cyan!85} 0.03 & \cellcolor{cyan!85} 0.12 & \cellcolor{cyan!85} 0.06 & \cellcolor{cyan!85} 0.01 & - & - & - & - \\

\bottomrule
\end{tabular}}
\end{table*}

\begin{table*}[h!]
\centering
\caption{\textcolor{black}{Comparison of $\hat{A}_{12}$ effect sizes for \pimodel. Table interpretation is the same as Table \ref{tab:RQ3_OpenVLA}}}
\label{tab:RQ3_pi0}
\resizebox{\textwidth}{!}{
\begin{tabular}{llcccccccccccc}
\toprule
& & \multicolumn{3}{c}{\textbf{Pick up}}& \multicolumn{3}{c}{\textbf{Move Near}}& \multicolumn{3}{c}{\textbf{Put in}}& \multicolumn{3}{c}{\textbf{Put on}}\\
\cmidrule{2-14}
& & High & Medium & Low & High & Medium & Low & High & Medium & Low & High & Medium & Low \\
\cmidrule{1-14}
\multirow{16}{*}{\textbf{Uncertainty Metrics}}&  TB-TP & \cellcolor{cyan!85} 0.08 & \cellcolor{cyan!85} 0.15 & \cellcolor{cyan!60} 0.33 & \cellcolor{cyan!85} 0.24 & 0.41 & 0.43 & 0.54 & 0.60 & 0.57 & \cellcolor{red!85} 0.80 & \cellcolor{red!30} 0.65 & 0.52 \\
 & TB-PCS & \cellcolor{cyan!85} 0.20 & \cellcolor{cyan!85} 0.22 & \cellcolor{cyan!30} 0.40 & \cellcolor{cyan!85} 0.22 & 0.34 & \cellcolor{cyan!30} 0.39 & 0.58 & \cellcolor{red!30} 0.62 & 0.59 & \cellcolor{red!85} 0.79 & 0.61 & 0.47 \\
  & TB-D & 0.50 & 0.50 & 0.50 & 0.50 & 0.50 & 0.50 & 0.50 & 0.50 & 0.50 & 0.50 & 0.50 & 0.50 \\
 & TB-E & \cellcolor{cyan!30} 0.34 & \cellcolor{cyan!30} 0.36 & \cellcolor{cyan!30} 0.42 & \cellcolor{cyan!85} 0.17 & 0.32 & \cellcolor{cyan!30} 0.38 & 0.58 & \cellcolor{red!30} 0.62 & 0.51 & 0.41 & 0.53 & \cellcolor{red!30} 0.63 \\
& A-PI$_{all}$ & \cellcolor{cyan!85} 0.10 & 0.40 & \cellcolor{red!60} 0.67 & \cellcolor{cyan!60} 0.27 & 0.45 & 0.47 & 0.44 & 0.54 & 0.42 & \cellcolor{cyan!60} 0.29 & 0.46 & \cellcolor{red!30} 0.61 \\
 & A-PI$_{o}$ & \cellcolor{cyan!85} 0.09 & \cellcolor{cyan!60} 0.27 & 0.52 & \cellcolor{cyan!85} 0.17 & \cellcolor{cyan!60} 0.27 & \cellcolor{cyan!30} 0.37 & \cellcolor{cyan!60} 0.33 & 0.42 & 0.44 & \cellcolor{cyan!85} 0.20 & \cellcolor{cyan!60} 0.30 & 0.45 \\
 & A-PI$_{p}$ & \cellcolor{cyan!85} 0.13 & \cellcolor{cyan!60} 0.28 & 0.56 & \cellcolor{cyan!85} 0.20 & 0.34 & \cellcolor{cyan!30} 0.37 & \cellcolor{cyan!60} 0.30 & 0.44 & 0.45 & \cellcolor{cyan!85} 0.24 & \cellcolor{cyan!60} 0.33 & 0.48 \\
& A-VI$_{all}$ & \cellcolor{cyan!85} 0.10 & 0.39 & \cellcolor{red!30} 0.66 & \cellcolor{cyan!60} 0.28 & 0.46 & 0.47 & 0.45 & 0.54 & 0.43 & \cellcolor{cyan!60} 0.29 & 0.46 & \cellcolor{red!30} 0.61 \\
 & A-VI$_{o}$ & \cellcolor{cyan!85} 0.08 & \cellcolor{cyan!85} 0.24 & 0.49 & \cellcolor{cyan!85} 0.15 & \cellcolor{cyan!85} 0.26 & \cellcolor{cyan!30} 0.36 & \cellcolor{cyan!30} 0.35 & 0.41 & 0.44 & \cellcolor{cyan!85} 0.20 & \cellcolor{cyan!60} 0.27 & 0.43 \\
 & A-VI$_{p}$ & \cellcolor{cyan!85} 0.14 & \cellcolor{cyan!85} 0.25 & 0.49 & \cellcolor{cyan!85} 0.17 & \cellcolor{cyan!60} 0.30 & \cellcolor{cyan!30} 0.35 & \cellcolor{cyan!60} 0.31 & 0.43 & 0.42 & \cellcolor{cyan!85} 0.23 & \cellcolor{cyan!60} 0.31 & 0.45 \\
 & A-AI$_{all}$ & \cellcolor{cyan!85} 0.10 & 0.40 & \cellcolor{red!30} 0.66 & \cellcolor{cyan!60} 0.29 & 0.47 & 0.47 & 0.44 & 0.55 & 0.43 & \cellcolor{cyan!60} 0.29 & 0.47 & \cellcolor{red!30} 0.62 \\
 & A-AI$_{o}$ & \cellcolor{cyan!85} 0.09 & \cellcolor{cyan!85} 0.24 & 0.48 & \cellcolor{cyan!85} 0.14 & \cellcolor{cyan!85} 0.26 & \cellcolor{cyan!30} 0.36 & \cellcolor{cyan!30} 0.36 & 0.42 & 0.43 & \cellcolor{cyan!85} 0.20 & \cellcolor{cyan!60} 0.27 & 0.43 \\
 & A-AI$_{p}$ & \cellcolor{cyan!85} 0.14 & \cellcolor{cyan!85} 0.25 & 0.47 & \cellcolor{cyan!85} 0.16 & \cellcolor{cyan!60} 0.29 & \cellcolor{cyan!30} 0.35 & \cellcolor{cyan!60} 0.32 & 0.44 & 0.43 & \cellcolor{cyan!85} 0.23 & \cellcolor{cyan!60} 0.30 & 0.44 \\
 & EV & \cellcolor{cyan!30} 0.40 & 0.47 & \cellcolor{red!30} 0.62 & \cellcolor{cyan!85} 0.25 & 0.40 & 0.41 & 0.52 & 0.55 & 0.41 & \cellcolor{cyan!85} 0.24 & 0.42 & 0.58 \\
\cmidrule{1-14}
\multirow{11}{*}{\textbf{Quality Metrics}}  & TCP-PI$_{all}$ & \cellcolor{cyan!85} 0.22 & 0.39 & \cellcolor{red!30} 0.59 & \cellcolor{cyan!85} 0.22 & 0.40 & 0.45 & \cellcolor{cyan!30} 0.38 & 0.49 & 0.51 & \cellcolor{cyan!60} 0.29 & 0.39 & 0.52 \\
 & TCP-PI$_{o}$ & \cellcolor{cyan!85} 0.23 & 0.40 & 0.56 & \cellcolor{cyan!85} 0.23 & 0.42 & 0.48 & \cellcolor{cyan!30} 0.38 & 0.48 & 0.49 & \cellcolor{cyan!60} 0.28 & 0.38 & 0.50 \\
 & TCP-PI$_{p}$ & \cellcolor{cyan!85} 0.22 & \cellcolor{cyan!30} 0.38 & 0.57 & \cellcolor{cyan!85} 0.26 & 0.40 & 0.43 & 0.44 & 0.56 & 0.57 & \cellcolor{cyan!30} 0.36 & 0.48 & 0.59 \\
& TCP-VI$_{all}$ & \cellcolor{cyan!85} 0.14 & \cellcolor{cyan!30} 0.34 & 0.57 & \cellcolor{cyan!85} 0.19 & 0.37 & 0.42 & \cellcolor{cyan!30} 0.35 & 0.44 & 0.47 & \cellcolor{cyan!85} 0.26 & \cellcolor{cyan!30} 0.35 & 0.50 \\
 & TCP-VI$_{o}$ & \cellcolor{cyan!85} 0.14 & \cellcolor{cyan!30} 0.35 & 0.54 & \cellcolor{cyan!85} 0.20 & 0.41 & 0.46 & \cellcolor{cyan!30} 0.37 & 0.45 & 0.48 & \cellcolor{cyan!85} 0.24 & \cellcolor{cyan!30} 0.34 & 0.48 \\
 & TCP-VI$_{p}$ & \cellcolor{cyan!85} 0.16 & \cellcolor{cyan!30} 0.35 & 0.56 & \cellcolor{cyan!85} 0.21 & 0.36 & \cellcolor{cyan!30} 0.39 & \cellcolor{cyan!30} 0.34 & 0.44 & 0.47 & \cellcolor{cyan!60} 0.32 & 0.42 & 0.54 \\
 & TCP-AI$_{all}$ & \cellcolor{cyan!85} 0.14 & \cellcolor{cyan!60} 0.32 & 0.52 & \cellcolor{cyan!85} 0.17 & 0.34 & 0.41 & \cellcolor{cyan!30} 0.36 & 0.44 & 0.47 & \cellcolor{cyan!85} 0.23 & \cellcolor{cyan!60} 0.31 & 0.48 \\
 & TCP-AI$_{o}$ & \cellcolor{cyan!85} 0.13 & \cellcolor{cyan!60} 0.33 & 0.51 & \cellcolor{cyan!85} 0.19 & 0.41 & 0.46 & \cellcolor{cyan!30} 0.38 & 0.45 & 0.48 & \cellcolor{cyan!85} 0.23 & \cellcolor{cyan!60} 0.31 & 0.48 \\
 & TCP-AI$_{p}$ & \cellcolor{cyan!85} 0.16 & \cellcolor{cyan!60} 0.32 & 0.51 & \cellcolor{cyan!85} 0.19 & 0.31 & \cellcolor{cyan!30} 0.37 & \cellcolor{cyan!60} 0.33 & 0.42 & 0.44 & \cellcolor{cyan!60} 0.28 & 0.37 & 0.51 \\
& TI & \cellcolor{cyan!85} 0.17 & \cellcolor{cyan!30} 0.35 & 0.54 & \cellcolor{cyan!85} 0.20 & 0.33 & \cellcolor{cyan!30} 0.39 & \cellcolor{cyan!60} 0.32 & 0.43 & 0.45 & \cellcolor{cyan!60} 0.31 & 0.40 & 0.53 \\
& OT & \cellcolor{cyan!85} 0.06 & \cellcolor{cyan!85} 0.14 & \cellcolor{cyan!60} 0.27 & \cellcolor{cyan!85} 0.06 & \cellcolor{cyan!85} 0.03 & \cellcolor{cyan!85} 0.13 & \cellcolor{cyan!85} 0.05 & \cellcolor{cyan!85} 0.12 & \cellcolor{cyan!85} 0.07 & \cellcolor{cyan!85} 0.15 & \cellcolor{cyan!85} 0.10 & \cellcolor{cyan!85} 0.14 \\

\bottomrule
\end{tabular}}
\end{table*}

\begin{table*}[h!]
\centering
\caption{\textcolor{black}{Comparison of $\hat{A}_{12}$ effect sizes for SpatialVLA. Table interpretation is the same as Table \ref{tab:RQ3_OpenVLA}}}
\label{tab:RQ3_spatialVLA}
\resizebox{\textwidth}{!}{
\begin{tabular}{llcccccccccccc}
\toprule
& & \multicolumn{3}{c}{\textbf{Pick up}}& \multicolumn{3}{c}{\textbf{Move Near}}& \multicolumn{3}{c}{\textbf{Put in}}& \multicolumn{3}{c}{\textbf{Put on}}\\
\cmidrule{2-14}
& & High & Medium & Low & High & Medium & Low & High & Medium & Low & High & Medium & Low \\

\cmidrule{1-14}
\multirow{16}{*}{\textbf{Uncertainty Metrics}} & TB-TP & \cellcolor{cyan!85} 0.11 & \cellcolor{cyan!60} 0.27 & 0.48 & \cellcolor{cyan!30} 0.35 & 0.56 & 0.56 & 0.40 & 0.53 & - & \cellcolor{cyan!60} 0.30 & \cellcolor{cyan!60} 0.27 & 0.47 \\
 & TB-PCS & \cellcolor{cyan!85} 0.12 & \cellcolor{cyan!60} 0.27 & 0.46 & \cellcolor{cyan!60} 0.33 & 0.53 & 0.55 & 0.41 & 0.54 & - & \cellcolor{cyan!60} 0.30 & \cellcolor{cyan!60} 0.28 & 0.48 \\
 & TB-D & \cellcolor{cyan!85} 0.10 & \cellcolor{cyan!60} 0.27 & 0.49 & \cellcolor{cyan!30} 0.34 & 0.57 & 0.58 & 0.40 & 0.54 & - & \cellcolor{cyan!60} 0.30 & \cellcolor{cyan!60} 0.27 & 0.47 \\
 & TB-E & \cellcolor{cyan!85} 0.10 & \cellcolor{cyan!60} 0.27 & 0.52 & \cellcolor{cyan!30} 0.35 & 0.59 & 0.58 & \cellcolor{cyan!30} 0.37 & 0.53 & - & \cellcolor{cyan!60} 0.29 & \cellcolor{cyan!60} 0.27 & 0.45 \\
 & A-PI$_{all}$ & \cellcolor{cyan!85} 0.16 & \cellcolor{cyan!30} 0.39 & \cellcolor{red!30} 0.62 & 0.52 & \cellcolor{red!60} 0.70 & \cellcolor{red!30} 0.65 & 0.53 & 0.62 & - & \cellcolor{cyan!30} 0.37 & 0.49 & 0.62 \\
 & A-PI$_{o}$ & \cellcolor{cyan!85} 0.20 & \cellcolor{cyan!30} 0.39 & \cellcolor{red!30} 0.61 & 0.45 & \cellcolor{red!30} 0.65 & \cellcolor{red!30} 0.63 & 0.43 & 0.55 & - & \cellcolor{cyan!30} 0.34 & \cellcolor{cyan!30} 0.35 & 0.49 \\
 & A-PI$_{p}$ & \cellcolor{cyan!85} 0.23 & 0.40 & \cellcolor{red!30} 0.64 & 0.44 & \cellcolor{red!30} 0.65 & 0.60 & 0.47 & 0.64 & - & 0.44 & 0.39 & 0.58 \\
& A-VI$_{all}$ & \cellcolor{cyan!85} 0.17 & \cellcolor{cyan!30} 0.39 & \cellcolor{red!30} 0.62 & 0.54 & \cellcolor{red!60} 0.71 & \cellcolor{red!30} 0.64 & 0.54 & 0.63 & - & \cellcolor{cyan!30} 0.37 & 0.49 & 0.62 \\
 & A-VI$_{o}$ & \cellcolor{cyan!85} 0.20 & \cellcolor{cyan!30} 0.39 & 0.60 & 0.43 & \cellcolor{red!30} 0.65 & \cellcolor{red!30} 0.62 & 0.41 & 0.54 & - & \cellcolor{cyan!30} 0.34 & \cellcolor{cyan!30} 0.34 & 0.48 \\
 & A-VI$_{p}$ & \cellcolor{cyan!85} 0.22 & 0.40 & \cellcolor{red!30} 0.62 & \cellcolor{cyan!30} 0.42 & \cellcolor{red!30} 0.64 & 0.60 & 0.44 & 0.62 & - & 0.44 & 0.37 & 0.56 \\
 & A-AI$_{all}$ & \cellcolor{cyan!85} 0.17 & \cellcolor{cyan!30} 0.39 & \cellcolor{red!30} 0.62 & 0.54 & \cellcolor{red!60} 0.72 & \cellcolor{red!30} 0.64 & 0.54 & 0.63 & - & \cellcolor{cyan!30} 0.38 & 0.50 & 0.63 \\
 & A-AI$_{o}$ & \cellcolor{cyan!85} 0.20 & 0.40 & 0.60 & 0.42 & \cellcolor{red!30} 0.65 & \cellcolor{red!30} 0.62 & 0.41 & 0.54 & - & \cellcolor{cyan!30} 0.34 & \cellcolor{cyan!30} 0.34 & 0.48 \\
 & A-AI$_{p}$ & \cellcolor{cyan!85} 0.22 & 0.41 & \cellcolor{red!30} 0.62 & \cellcolor{cyan!30} 0.41 & \cellcolor{red!30} 0.65 & 0.59 & 0.44 & 0.61 & - & 0.45 & 0.37 & 0.57 \\
 & EV & 0.50 & 0.50 & 0.50 & 0.48 & 0.48 & 0.48 & 0.55 & 0.40 & - & 0.60 & \cellcolor{cyan!30} 0.36 & 0.52 \\
\cmidrule{1-14}
\multirow{11}{*}{\textbf{Quality Metrics}} & TCP-PI$_{all}$ & 0.48 & \cellcolor{cyan!60} 0.33 & 0.57 & \cellcolor{red!60} 0.68 & \cellcolor{red!60} 0.68 & \cellcolor{red!60} 0.69 & 0.55 & 0.63 & - & 0.45 & 0.47 & 0.59 \\
 & TCP-PI$_{o}$ & \cellcolor{red!60} 0.68 & 0.50 & 0.51 & \cellcolor{red!60} 0.68 & 0.64 & \cellcolor{red!30} 0.66 & 0.52 & 0.58 & - & \cellcolor{cyan!30} 0.39 & 0.42 & 0.54 \\
 & TCP-PI$_{p}$ & \cellcolor{cyan!60} 0.30 & \cellcolor{cyan!60} 0.31 & 0.60 & \cellcolor{red!30} 0.64 & \cellcolor{red!60} 0.68 & \cellcolor{red!60} 0.68 & \cellcolor{red!85} 0.76 & \cellcolor{red!85} 0.82 & - & 0.54 & 0.54 & \cellcolor{red!60} 0.67 \\
 & TCP-VI$_{all}$ & \cellcolor{cyan!85} 0.21 & 0.40 & \cellcolor{red!30} 0.63 & 0.47 & \cellcolor{red!30} 0.66 & \cellcolor{red!30} 0.66 & 0.44 & 0.61 & - & \cellcolor{cyan!30} 0.35 & \cellcolor{cyan!30} 0.37 & 0.51 \\
 & TCP-VI$_{o}$ & \cellcolor{cyan!85} 0.22 & 0.43 & 0.60 & 0.47 & \cellcolor{red!60} 0.67 & \cellcolor{red!60} 0.68 & 0.44 & 0.59 & - & \cellcolor{cyan!60} 0.32 & \cellcolor{cyan!30} 0.35 & 0.48 \\
 & TCP-VI$_{p}$ & \cellcolor{cyan!85} 0.22 & \cellcolor{cyan!30} 0.38 & \cellcolor{red!30} 0.64 & 0.49 & \cellcolor{red!60} 0.68 & \cellcolor{red!30} 0.62 & 0.52 & \cellcolor{red!60} 0.73 & - & 0.44 & 0.42 & 0.58 \\
 & TCP-AI$_{all}$ & \cellcolor{cyan!85} 0.19 & \cellcolor{cyan!30} 0.39 & 0.60 & 0.44 & \cellcolor{red!30} 0.66 & \cellcolor{red!30} 0.65 & 0.42 & 0.59 & - & \cellcolor{cyan!60} 0.33 & \cellcolor{cyan!60} 0.33 & 0.47 \\
 & TCP-AI$_{o}$ & \cellcolor{cyan!85} 0.21 & 0.43 & 0.58 & 0.45 & \cellcolor{red!60} 0.67 & \cellcolor{red!60} 0.67 & 0.42 & 0.59 & - & \cellcolor{cyan!60} 0.31 & \cellcolor{cyan!60} 0.33 & 0.45 \\
 & TCP-AI$_{p}$ & \cellcolor{cyan!85} 0.20 & \cellcolor{cyan!30} 0.37 & 0.61 & 0.45 & \cellcolor{red!60} 0.67 & \cellcolor{red!30} 0.63 & 0.46 & \cellcolor{red!60} 0.68 & - & 0.42 & 0.37 & 0.55 \\
 & TI & \cellcolor{cyan!85} 0.22 & \cellcolor{cyan!30} 0.38 & \cellcolor{red!30} 0.65 & 0.47 & \cellcolor{red!30} 0.65 & \cellcolor{red!30} 0.63 & 0.52 & \cellcolor{red!60} 0.71 & - & 0.43 & 0.40 & 0.57 \\
& OT & \cellcolor{cyan!85} 0.05 & \cellcolor{cyan!85} 0.23 & \cellcolor{cyan!60} 0.32 & \cellcolor{cyan!30} 0.40 & \cellcolor{cyan!85} 0.23 & \cellcolor{cyan!85} 0.22 & \cellcolor{cyan!85} 0.09 & \cellcolor{cyan!85} 0.11 & - & \cellcolor{cyan!85} 0.23 & \cellcolor{cyan!85} 0.14 & \cellcolor{cyan!85} 0.13 \\

\bottomrule
\end{tabular}}
\end{table*}

Among our suite of uncertainty and quality metrics, OT consistently emerged as the single most powerful discriminator metric between successful and failing tasks. OT achieved statistically significant and large $\hat{A}_{12}$ effect sizes across all models and tasks. Only for the ``\textit{Pick up}'' task and low-quality executions in the OpenVLA model, this metric did not show statistical significance with respect to the unsuccessful tasks. This highlights the utility of the OT metric as a model‑agnostic indicator of failures in VLA‑enabled robotic systems.

\revision{This strong discrimination of OT can be explained by the definition of OT itself. Since it directly measures the optimal trajectory to the task target, successful executions inherently yield trajectories that converge to the goal, resulting in distances close to zero at the final stage. In contrast, failed executions terminate with the end-effector remaining at non-zero distances from the target. As a consequence, OT exhibits a direct relationship with task completion quality, which naturally leads to a clear separation between successful and unsuccessful tasks.}

\begin{FindingBox}
OT serves as a consistent, model‑agnostic indicator of successful and unsuccessful tasks.
\end{FindingBox}

When considering the tasks labeled as high-quality, apart from OT, several other metrics exhibited task- and model-specific discrimination capabilities.
In the \textit{``Pick up''} task, all metrics (both quality and uncertainty) demonstrated discrimination capabilities between high-quality and failing tasks, with the following exceptions: In OpenVLA, there is no metric that showed discrimination capabilities; for the \pimodel, TB-D 
did not show statistical significance;  in SpatialVLA, EV did not show statistical significance.

Similarly, in the \textit{``Move near''} task for the \pimodel{} model, all metrics except EV reliably distinguished high-quality tasks from failing ones, and SpatialVLA’s TB‑derived metrics (TB‑TP, TB‑PCS, TB‑D, TB‑E) demonstrated similar performance. For OpenVLA in the \textit{``Put in''} task, these token-based metrics proved to be useful for distinguishing high-quality and failing tasks. Furthermore, for the \pimodel{} model we found that every quality metric except the TPC-PI maintained significant discrimination across every task between the high-quality successful tasks and the failing ones.

With these results we foresee two prominent directions. On the one hand, the quality metrics can be reliably used at design-time to assess whether a task is being successful or not, without relying on symbolic oracles, by establishing certain thresholds. On the other hand, uncertainty metrics can be reliably used at runtime for monitoring and ensuring that the quality of the task is high-enough, detecting unsuccessful tasks in an automated manner. These applications are especially important when these tasks are in high-stake and critical applications where high-quality is paramount.

\begin{FindingBox}
Apart from OT, except for the OpenVLA model, most of the uncertainty and quality metrics showed promising results at distinguishing high-quality task executions with failing ones.

\end{FindingBox}

\revision{These results also have important practical implications. In particular, they suggested that uncertainty and quality metrics can be effectively used to distinguish high-quality executions from failing ones, but their reliability was strongly model-dependent. Metrics such as EV, for instance, losed their interpretability when the underlying policy was largely deterministic or exhibited limited stochasticity, as observed in SpatialVLA and parts of OpenVLA. In such cases, EV no longer reflected meaningful uncertainty, but rather collapsed to nearly constant or uninformative values.}

\revision{The weaker performance observed in OpenVLA further highlighted the dependence of metric discriminability on the underlying policy quality. In this model, the relatively low number of successful high-quality executions, combined with behavioral similarity between successes and failures, leaded to substantial overlap in metric distributions. As a result, many failing trajectories exhibited behavior similar to successful ones (i.e., the failing ones frequently didn't move, thus generating near-zero values on several metrics), making separation more challenging as shown by the distribution of each metric in Figure~\ref{fig:RQ2-openvla}. This indicates that the effectiveness of uncertainty and quality metrics is not solely an intrinsic property of the metrics themselves, but also depends on the structure and quality of the policy being evaluated. As models become more accurate and their failure modes more structured, successful and failing trajectories become more separable, which in turn improves the discriminative power of the proposed metrics. Conversely, in weaker or less consistent policies, metric distributions tend to overlap, reducing discrimination capabilities.}

\revision{Overall, this suggests that these metrics are most useful in scenarios where the policy exhibits a minimum level of structured and meaningful behavior, reinforcing their role as complementary tools for evaluating increasingly capable VLA-enabled robots.}

\begin{FindingBox}
\revision{The effectiveness of uncertainty and quality metrics for discriminating between successful and failing executions depends on the underlying model. Metrics such as EV may become uninformative for near-deterministic policies, and overall separability improves as the policy becomes more accurate and exhibits more structured failure modes.}
\end{FindingBox}

With respect to the discrimination between medium-quality and failing tasks, for SpatialVLA all metrics except EV showed discrimination capabilities in the \textit{``Pick up''} task. Likewise, for the \pimodel{} model all quality metrics as well as some uncertainty metrics, i.e., TB-TP, TB-PCS and TB-E, showed discrimination capabilities. Moreover, TB-based metrics showed statistical significance when discriminating between medium-quality tasks and unsuccessful tasks for OpenVLA model in the \textit{``Put  In''} task, as well as for the SpatialVLA model in \textit{``Put on''} task with at least medium effect size of $\hat{A}_{12}$. 
However, for SpatialVLA in \textit{``Move near''} and \textit{``Put in''} task all quality metrics except OT did not show capacity for discriminating between medium-quality tasks and unsuccessful tasks. In most of the failed tasks, we observed that the robot began executing a task but quickly became blocked or collided with the environment, entering a \textit{blocking} state early in the process. In contrast, for successful tasks, this \textit{blocking} state typically occurred only after the task had already been completed, much later in the execution timeline compared to the failing tasks. As a result, the average values of these metrics tend to be higher for medium-quality successful tasks compared to failed ones.

As for low-quality tasks, apart from OT, with the exception of some metrics in the \pimodel{} model for the  \textit{``Pick up''} and the \textit{``Move near''} task, the metrics did not show discrimination capabilities between low-quality and failing tasks. This suggests that many of the low-quality, yet successful tasks, are quite close to those non-successful tasks when considering our defined metrics.

\begin{FindingBox}
While our metrics revealed statistically significant differences that were specific to both model and task, most, except for OT, failed to clearly distinguish between medium- and low-quality successful tasks and outright failures. This suggests that as task quality declines, outcomes increasingly resemble failures, reinforcing our conclusion that success rate alone is insufficient for evaluating the quality of VLA models.
\end{FindingBox}

Interestingly, some failing executions showed favorable results when applying our metrics. For instance, we showed cases with low instability, potentially due to the VLA model not being able to localize or detect a target object. In such situations, the robot did not move at all, which showed non acceleration or velocities in its end effector. In these particular cases, action‑ and motion‑based metrics (e.g., A‑PI, A‑VI, A‑AI, TI, TCP‑PI, TCP‑VI, TCP‑AI) showed minimal deviation and instability, values that under normal circumstances indicate high‑quality behavior; instead, in these specific cases, low values merely reflect a lack of motion rather than successful task execution. These situations could eventually be easily detected by establishing a minimal threshold that measures some motion in the robot.

\begin{FindingBox}
Some unsuccessful tasks showed low uncertainty or high quality, as no instability was detected. This, however, was due to the robot being static. Such situations can be easily detected by establishing a threshold that measures minimal robot motion.

\end{FindingBox}

\revision{Overall, the results showed that most uncertainty and quality metrics demonstrated strong discrimination capabilities between high-quality successful and failing tasks across multiple models and tasks. These findings suggest that such metrics can serve as effective indicators of execution quality, enabling the automated identification of unsuccessful or degraded behaviors without relying exclusively on symbolic task oracles. Consequently, these metrics show strong potential for supporting software engineering tasks such as testing, and runtime monitoring of robotic task executions, especially in safety-critical scenarios where maintaining consistently high-quality behavior is essential. }

\begin{FindingBox}
\revision{The ability of uncertainty and quality metrics to discriminate between high-quality and failing executions suggests that they can effectively support testing, and runtime monitoring of VLA-enabled robotic systems, particularly in critical scenarios where high-quality task execution is required.}
\end{FindingBox}

\subsection{RQ4 - Overhead}
RQ4 assessed the \revision{computational overhead and practical applicability of each metric when used for testing and runtime monitoring of VLA-enabled robots. This is particularly important for uncertainty metrics intended for online deployment (e.g., for self-healing task or real-time failure detection during task execution).} 

Table~\ref{tab:exec_time_summary} reports the overall inference time and corresponding overhead for each of the metric. Note that the overhead of the TB-TP, TB-PCS, TB-D, and TB-D metrics was measured together because it is mainly caused by the process of obtaining token probabilities. In addition, we grouped A-PI, A-VI, and A-AI under AI, as computing A-AI requires prior computation of A-PI and A-VI. Similarly, TCP-AI, which depends on TCP-PI and TCP-VI, was grouped together with them under TCP.

\begin{table}[h!]
\centering
\caption{Mean ($m$) and standard deviation ($\sigma$) of inference time (in seconds) for each model and execution overhead (in seconds) for each model and metric.}
\label{tab:exec_time_summary}

\resizebox{0.485\textwidth}{!}{
\begin{tabular}{lrrrrrr}
\toprule
 & \multicolumn{2}{c}{\textbf{OpenVLA}} & \multicolumn{2}{c}{\pimodel} & \multicolumn{2}{c}{\textbf{SpatialVLA}} \\
 \cmidrule{2-7}
 & \multicolumn{1}{c}{$m$} & \multicolumn{1}{c}{$\sigma$} & \multicolumn{1}{c}{$m$} & \multicolumn{1}{c}{$\sigma$} & \multicolumn{1}{c}{$m$} & \multicolumn{1}{c}{$\sigma$} \\
 \cmidrule{1-7}
 \textbf{Inference} & 0.363107 & 0.029336 & 0.489265 & 0.025283 & 0.751276 & 0.045674 \\
\cmidrule{1-7}
\textbf{TB} & 0.012696 & 0.464 & 0.069965 & 0.003203 & 0.153026 & 0.009194 \\
\textbf{AI} & 0.114 & 0.007 & 0.115 & 0.008 & 0.106 & 0.011 \\
\textbf{EV} & 1.057714 & 0.070556 & 1.916454 & 0.076461 & 2.937309 & 0.430298 \\
\cmidrule{1-7}
\textbf{TCP} & 0.114 & 0.007 & 0.115 & 0.008 & 0.106 & 0.011 \\

\textbf{TI} & 0.283 & 0.034 & 0.421 & 0.033 & 0.239 & 0.040 \\
\textbf{OT} & 0.117 & 0.019 & 0.135 & 0.019 & 0.142 & 0.045 \\
\bottomrule
\end{tabular}}
\end{table}

As expected, the EV metric showed the highest overhead since a VLA model needs to be instantiated multiple times to obtain and compare the results of multiple inferences. While this is feasible in simulation, the overhead is too high for practical use, especially in scenarios where models need to be executed frequently to effectively control robot actions. Combined with RQ2b results for the EV metric (which showed no positive correlation in most cases), makes this metric unsuitable for measuring the uncertainty of VLA-enabled robots.

The remaining metrics do not have significant overhead, with the TB metrics producing the highest overhead, \revision{due to the token probability extraction, yet the added latency remains low enough for practical integration into robotic control loops. The AI- and TCP-based metrics, as well as TI and OT, add negligible computational cost.}

\begin{FindingBox}
\revision{The EV metric adds excessively high inference overhead due to the need for multiple model executions, making it impractical for real-world. In contrast, the remaining metrics introduce only low overhead and are suitable for practical use.}
\end{FindingBox}

\revision{However, practical applicability extends beyond raw computational cost. While most metrics are lightweight, OT relies on the assumption that effective executions consistently reduce the distance to the goal over time. This assumption holds well in simple, obstacle-free simulated environments, but becomes problematic in realistic scenarios involving detours, obstacle avoidance, or dynamic trajectory adjustments. In such cases, continuously measuring the optimal trajectory distance is often infeasible without complete knowledge of the environment due to high environmental uncertainty. Consequently, like EV, OT is primarily suitable for offline simulation-based evaluation rather than online deployment on physical robots.} 

\begin{FindingBox}
\revision{Most proposed metrics (TB, AI, TCP, and TI) are lightweight enough for practical integration into the testing and runtime monitoring of VLA-enabled robots. However, the OT metric is largely limited to simulation environments due to the difficulty of measuring optimal trajectories in the real world.}
\end{FindingBox}

\section{Threats to Validity}
\label{sec:threats}


Some of our metrics required certain parameters, which may lead to potential \textbf{internal validity} threats. For instance, the EV requires multiple inferences; we limited the number to 4, constrained by the hardware capabilities of our infrastructure. Nevertheless, this configuration demonstrated stable results across evaluations. On the other hand, we established a time window of 8 steps in those metrics that required temporal differences for their computations. This value was selected as we empirically observed that lower values led to less stable estimations, while 8 steps provided the best trade-off between temporal context and computational cost. Another \textbf{internal validity} threat could relate to model selection and training heterogeneity. We mitigated such a threat by using models that were fine-tuned with the same dataset across different tasks. \revision{Another potential internal validity threat concerns the use of two different servers in our experimental infrastructure, as hardware differences could affect execution times and performance-related measurements. To mitigate this threat, all experiments associated with a given VLA model were executed exclusively on the same server, ensuring that all measurements and comparisons for that model were obtained under identical hardware conditions. Furthermore, our research questions do not rely on direct comparisons of raw metric values across different VLA models. Instead, the analyses are conducted per metric using the aggregated results from the three VLA models in our experimental setup, thereby reducing the potential impact of server-related differences on the validity of our conclusions.}

An \textbf{external validity threat} in our evaluation relates to the applicability of our approach beyond our specific case studies. Our evaluation spans the four primary tasks commonly addressed in the VLA model literature~\cite{wang2025vlatest, o2024open} and includes three state-of-the-art models with diverse architectural designs. However, generalizability is still constrained by the limited diversity of robotic platforms (two systems). \revision{We present our reported correlations and discrimination results primarily as in-domain findings for the three evaluated VLA models, four manipulation tasks, and two simulated robot embodiments. While these metrics are likely to generalize to similar VLA architectures and comparable manipulation settings, they are not intended for direct extrapolation to substantially different models, tasks, physical platforms, or real-world environments without additional validation.} To mitigate such a threat, we used a total of 500 diverse scene evaluation sets per task.

Our task quality was measured by involving domain experts in labeling the quality level of tasks, which may lead to a \textbf{conclusion validity threat}. To mitigate such threat, we involved a total of 3 experienced domain experts and established an adequate process to detect disagreement, with effective conflict resolution for those tasks labeled differently. Moreover, we developed a web-based system for easy tagging of successful tasks, distributed the labeling across a total of 15 days, and limited the number of tasks to label to a maximum of 160 tasks. This process showed high agreement between the first two labelers, showing an inter-rater agreement of 85\% according to Cohen's kappa.

\section{Related Work}
\label{sec:relatedwork}

\subsection{Uncertainty in DL Models and Foundation Models}

Deep learning (DL) models are known to achieve state-of-the-art performance in a wide range of tasks, while they often exhibit significant challenges in terms of uncertainty and reliability~\cite{abdar2021review}, particularly in safety-critical or high-stakes applications. Quantifying the uncertainty of DL models provides insights about the reliability and trustworthiness of the model predictions. To capture uncertainties, various approaches have been proposed~\cite{tran2019bayesian,gal2016dropout,yelleni2024monte,lakshminarayanan2017simple}. Among them, Bayesian Neural Networks (BNNs) offer a probabilistic framework for modeling uncertainty by applying Bayesian inference to DL models~\cite{tran2019bayesian}. 
However, BNNs are often computationally expensive and challenging to implement in practice due to the complexity of posterior inference. As a practical Bayesian approximation, Monte-Carlo Dropout (MC-Dropout) is a widely used uncertainty quantification (UQ) method~\cite{gal2016dropout}, which significantly reduces the computational cost compared to full Bayesian inference. 
Another prominent approach is Deep Ensembles (DE)~\cite{lakshminarayanan2017simple}, which involves training multiple neural networks independently with different random initializations. The variance among the ensemble models' predictions serves as a measure of uncertainty.

The above UQ methods have been widely applied to DL models in robotics and cyber-physical systems. For example, Xu et al.~\cite{xu2024pretrain} proposed an uncertainty-aware transfer learning method to evolve digital twins of CPSs, where Bayesian and ensemble methods are studied. Catak et al.~\cite{catak2022uncertainty} designed a prediction validator based on the MC-Dropout method. As for computer vision tasks, Feng et al.~\cite{feng2018towards} captured uncertainties using MC-Dropout in 3D vehicle detection models to improve vehicle detection performance and assure autonomous vehicle safety. Kendall and Gal~\cite{kendall2017uncertainties} investigated various uncertainty types for computer vision tasks, including semantic segmentation and depth regression. In robotics, Lu et al.\cite{10988977} evaluated the uncertainty and robustness of DL-based sticker detection software integrated into robotic arms and provided model selection guidelines based on the evaluation results. 
\revision{Beyond perception, uncertainty is also a central concern in robotic control, planning, and manipulation, where decisions are made under sensing noise, actuation error, and partial observability. Various studies incorporate uncertainty into the decision-making process. For example, Curtis et al.~\cite{curtis2024partially} proposed a risk-aware task-and-motion planning method under partial observability. Okada et al.~\cite{kuo2021uncertainty} designed an uncertainty-aware model-based control method for contact-rich manipulation. Other studies use uncertainty to decide when to act, such as calibrated confidence for selective execution and human hand-off~\cite{liang2024introspective,gaus2026confidence}. Different from these works, we do not design a planner or control policy, but focus on quantifying the predictive uncertainty of VLA-enabled robotics, which can be further used by such robotic decision-making policies.}

\revision{Capturing uncertainty in DL also requires the design of effective metrics, which are typically designed according to the tasks. For classification tasks, uncertainty is usually derived from softmax probabilities of class predictions. Commonly used metrics include Variation Ratio~\cite{freeman1965elementary}, Entropy~\cite{shannon1948mathematical}, Mutual Information~\cite{shannon1948mathematical}, Max Probability~\cite{10.1145/3417330}, PCS~\cite{scheffer2001active}, and DeepGini~\cite{feng2020deepgini}. For regression tasks, Total Variance~\cite{feng2018towards} is commonly employed to quantify predictive uncertainty. For object detection tasks, Prediction Surface Uncertainty~\cite{catak2021prediction} quantifies uncertainty as the convex-hull area spanned by the bounding-box predictions from multiple MC-Dropout passes, where a larger surface indicates higher localization uncertainty.}

Recent advances in foundation models (FM) have significantly expanded the capabilities of DL systems, while they also inherit uncertainty and reliability challenges from traditional DL models~\cite{shorinwa2025survey,huang2023look,10.1145/3641289,zhou2024larger}. 
\revision{However, existing UQ methods are not directly applicable to the FM context due to practical constraints such as computational costs and their pretrained nature. For instance, DE requires training multiple models from scratch, which is almost infeasible for large-scale FMs due to their enormous training costs and resource requirements.}
By adapting traditional DL epistemic uncertainty estimation techniques, Felicioni et al.~\cite{felicioni2024importance} studied the role of uncertainty in LLMs' decision-making with natural language as input.  
To understand uncertainty in pretrained LLMs, Xiao et al.~\cite{xiao2022uncertainty} conducted a large-scale empirical analysis to study a wide range of settings for LLMs, including the selection uncertainty quantifier. 
To better understand existing UQ approaches in the FM context, Huang et al.~\cite{huang2023look} conducted an exploratory study on the uncertainty assessment of LLMs, covering 12 uncertainty estimation \revision{metrics calculated based on token probabilities, such as Max Probability, Entropy, and Variation Ratio}.
Catak and Kuzlu~\cite{catak2024uncertainty} proposed a novel geometric approach to quantify the uncertainty in LLMs' responses \revision{and calculate the convex hull as the uncertainty metric.} 
To measure the trustworthiness of natural language responses generated by LLMs, Lin et al.~\cite{lin2024generatingconfidenceuncertaintyquantification} designed a set of metrics to capture the uncertainty and confidence of the input data and responses. 
Uncertainty estimation has been applied as a way to enhance the performance of FMs. For instance, Ji et al.~\cite{ji2023map} investigated uncertainty modeling in multimodal pretrained vision-language models and proposed a new module, Probability Distribution Encoder, which models uncertainty in multimodality as Gaussian distributions. Chen et al.~\cite{chen2024unveiling} investigated the correlation between uncertainty calibration and the performance of multimodal LLMs, and based on the evaluation results, several advanced multimodal LLM calibration techniques are proposed.

Recent advances have given rise to a new class of FMs known as VLA models, which integrate visual perception, natural language understanding, and action generation. Despite their growing importance, uncertainty quantification remains largely unexplored in the context of VLA models. This gap poses significant risks in high-stakes or safety-critical applications of VLAs. Therefore, in this work, we aim to design and adapt various UQ methods to the VLA context to assess their confidence and quality in robotic control. \revision{Specifically, we utilize the nondeterministic nature of VLA models and adapt four widely used UQ metrics (i.e., Entropy, Max Probability, PCS, and DeepGini) to the VLA context, calculated based on the token probabilities of the generated action sequences. In addition, we design novel output-based metrics that quantify uncertainty directly from the variability of the predicted actions across multiple stochastic inferences.}


\subsection{Quality Assessment in VLA-Enabled robots} 
Traditionally, quality assessment in robotics has relied on task-specific metrics that quantify performance in controlled environments~\cite{bohg2013data,mahler2017dex,lavalle2006planning,kavraki2002probabilistic}. For instance, navigation and path planning are commonly assessed using path length, execution time, and energy consumption~\cite{lavalle2006planning,kavraki2002probabilistic}. Similarly,  manipulation and grasping performance are assessed on success rates, completion times, and grasp reliability, where the success rate denotes the probability of a successful grasp~\cite{bohg2013data,mahler2017dex}. In parallel to robotics, computer vision and Natural Language Processing have established domain benchmarks to assess the quality of the system, such as ImageNet~\cite{deng2009imagenet} and Coco~\cite{lin2014microsoft} in vision and Bleu~\cite{papineni2002bleu} and Rouge~\cite{lin2004rouge} in language. Despite the utility of these single-modal metrics, they fall short when applied to complex multi-modal systems that integrate perception, language understanding, and action execution.

With the rise of intelligent agents for robotics, such as diffusion policies and VLA models, the community started to explore evaluation methods~\cite{wang2024ladev, wang2025vlatest} and benchmarks~\cite{li24simpler, o2024open, gong2023arnold, gu2023maniskill2, li2023behavior, makoviychuk2021isaac, mees2022calvin, zhu2020robosuite, liu2023libero} tailored to these models. Many of these benchmarks were primarily designed for simulation-based robotics, offering diverse scenes for a wide variety of actions, such as manipulation and navigation. They typically focus on scene setup and task completion, without capturing the full spectrum of capabilities required for VLA models, such as reasoning quality and task performance quality. For instance, in the benchmarks used to evaluate PaLM-E~\cite{driess2023palm} and RT-2~\cite{zitkovich2023rt}, the authors use an end-to-end evaluation that focuses on the final task performance of embodied agents in complex, multi-modal scenarios. In the case of PaLM-E, the model's evaluation emphasizes its ability to follow high-level natural language instructions to complete robotic manipulation and navigation tasks in real-world scenes. Similarly, RT-2 is evaluated by assessing its performance on a diverse suite of robotic tasks that require grounded visual reasoning and language comprehension. Both evaluation approaches mark a shift from traditional modular evaluation to integrated, outcome-oriented assessments. However, they still primarily measure end-task success rate without capturing fine-grained indicators of reasoning quality, cross-modal grounding, or real-time adaptability. 

Orthogonal to these approaches, Wang et al. introduced VLATest~\cite{wang2025vlatest}, a fuzzing framework designed to generate robotic manipulation scenes for testing VLA models, which also uses a simple oracle to assess the correctness of the task completion. They also proposed LADEV~\cite{wang2024ladev}, a platform that automatically constructs test environments from natural language descriptions, further enhanced by a paraphrasing mechanism to produce diverse task instruction variants. While these approaches represent a step toward a more systematic evaluation, they focused primarily on final outcomes, such as task completion rate and total execution time, similar to traditional robotics evaluations. In contrast to these existing approaches, our approach offers a comprehensive performance evaluation of the entire VLA-enabled robot, not only focusing on task completion, but encompassing the integration and quality of perception, language, reasoning, and execution.

\section{Conclusions and Future Work}
\label{sec:conclusion}

Visual Language Action (VLA) models are the next generation of AI-based control techniques for robotic manipulation tasks as well as other Cyber-Physical Systems. Yet, up to now, no one has in-depthly explored their quality and whether this can be measured through a set of formal quality and uncertainty metrics. In this paper, we propose different quality and uncertainty metrics for VLA-enabled robotic systems. We critically analyze the current evaluation frameworks, highlighting significant shortcomings in prevailing symbolic oracles to discriminate between successful and unsuccessful tasks. Our comprehensive manual analysis of 908 successful task executions across three leading VLA models revealed notable disparities in task performance quality, particularly underscoring the inadequacy of relying solely on success rates. The introduction and evaluation of eight uncertainty metrics and five quality metrics further illustrated their utility, as several metrics demonstrated moderate to high correlation with expert evaluations. Among these, Action Velocity Instability (A-VI), Action Acceleration Instability (A-AI), and Trajectory Velocity Instability (TCP-VI) stood out as particularly adequate indicators for assessing uncertainty and quality in this context. Additionally, Optimal Trajectory (OT) showed to be the best metric to distinguish between successful and failing tasks. Our findings emphasize the urgent need for standardized VLA evaluation and benchmarking frameworks and suggest promising avenues for enhancing real-time monitoring and adaptive performance improvements in robotic systems.

\revision{Our work also opens several promising avenues for future research. First, the metrics used in this work provide a foundation for the development of next-generation test oracles that move beyond binary correctness and incorporate behavioral quality during task execution. Such oracles could enable a more in-depth assessment of robotic systems by capturing properties such as stability, confidence, and execution smoothness. Second, the metrics could serve as objective functions in search-based testing, fuzzing, and test generation techniques. Rather than maximizing only failure rates, future approaches could systematically generate scenarios that maximize uncertainty or degrade execution quality, thereby exposing subtle weaknesses that remain undetected by traditional correctness-based evaluations. Third, several of the uncertainty metrics are suitable for runtime monitoring. Future work could investigate their integration into runtime decision-making frameworks that detect degraded behavior, trigger recovery strategies, or request human intervention when uncertainty exceeds acceptable thresholds. Such capabilities are particularly relevant for safety-critical applications of VLA-enabled robots. Fourth, the proposed metrics may provide the basis for behavioral test adequacy criteria and benchmarking methodologies. Existing evaluations largely focus on task completion rates, whereas our results suggest that uncertainty and quality metrics can capture important behavioral characteristics that are not reflected in binary outcomes. Future research could explore how these metrics can be incorporated into coverage criteria, benchmarking frameworks, and model comparison methodologies. Finally, our findings highlight several limitations that motivate further investigation. Some metrics impose substantial computational overhead, while others rely on information that may not be readily available in real-world deployments. Addressing these challenges will require research on lightweight metrics and techniques that bridge the simulation-to-reality gap, ultimately enabling the practical adoption of uncertainty- and quality-aware testing approaches in real-world robots. }

\section*{Acknowledgments}
Aitor Arrieta and Pablo Valle are part of the Systems and Software Engineering group of Mondragon Unibertsitatea (IT1519-22), supported by the Department of Education, Universities and Research of the Basque Country. Pablo Valle is supported by the Pre-doctoral Program for the Formation of Non-Doctoral Research Staff of the Education Department of the Basque Government (Grant n. PRE\_2024\_1\_0014). Aitor Arrieta is supported by the Spanish Ministry of Science, Innovation and Universities (project PID2023-152979OA-I00), funded by MCIU /AEI /10.13039/501100011033 / FEDER, UE. Chengjie Lu and Shaukat Ali are supported by the RoboSapiens project funded by the European Commission’s Horizon Europe programme under grant agreement number 101133807 and the Co-evolver project (No. 286898/F20) funded by the Research Council of Norway.

\ifCLASSOPTIONcaptionsoff
  \newpage
\fi

\bibliographystyle{ieeetr}
\bibliography{bibliography}

\appendices
\section{Implementation guidelines}

In this section, we provide relevant information about the architecture of the models used in our evaluation and the implementation details of the token-based uncertainty metrics, since these metrics have been implemented ad-hoc for each of the models. All the implementation scripts can be found in our Github repository~\cite{pvalle_VLA-UQ}.

\subsection{Models Architecture and Details} \label{app:ModelArch}

\begin{figure}[h]
    \centering
    
    \includegraphics[width=0.485\textwidth]{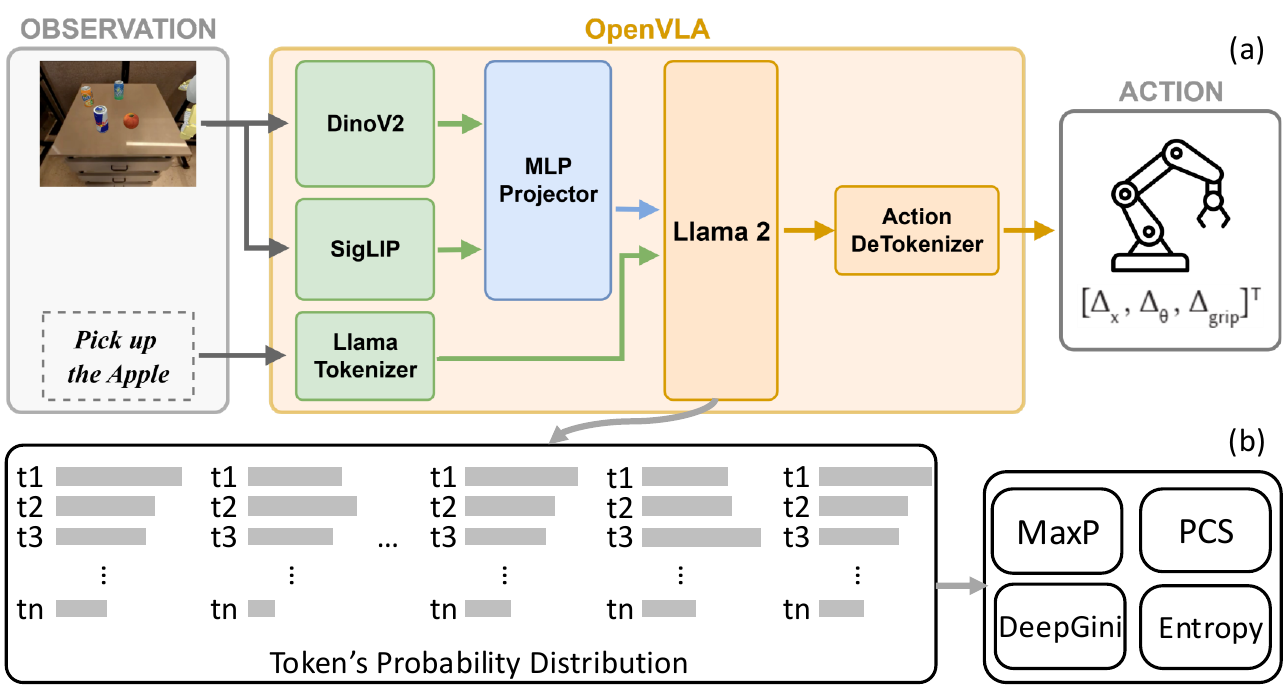}
    \caption{Architecture of OpenVLA}
    \label{fig:Arch_Openvla}
\end{figure}

\textbf{OpenVLA}~\cite{kim2024openvla} is a 7B-parameter VLA pretrained on 970k robot demonstrations from the Open X-Embodiment dataset~\cite{o2024open}. The architecture of OpenVLA consists of three main parts (See Figure~\ref{fig:Arch_Openvla} (a)): (1) a visual encoder that maps the image inputs to the image patch embeddings; (2) a projector that takes the output embeddings of the visual encoder and maps them into the input space of a language model; and (3) a large language model (LLM) backbone.  Particularly, OpenVLA is built on the Pirsmatic-7b VLM~\cite{karamcheti2024prismatic}, which follows the standard architecture described above, with a 600M-parameter visual encoder, a small 2-layer MLP projector, and a 7B-parameter Llama 2 language model backbone~\cite{touvron2023llama}. It is noteworthy that Prismatic uses a two-part visual encoder, consisting of pretrained SigLIP~\cite{zhai2023sigmoid} and DinoV2~\cite{oquab2023dinov2} models. Input images are passed separately through both encoders and the resulting feature vectors are concatenated.

\begin{figure}[h]
    \centering
    
    \includegraphics[width=0.485\textwidth]{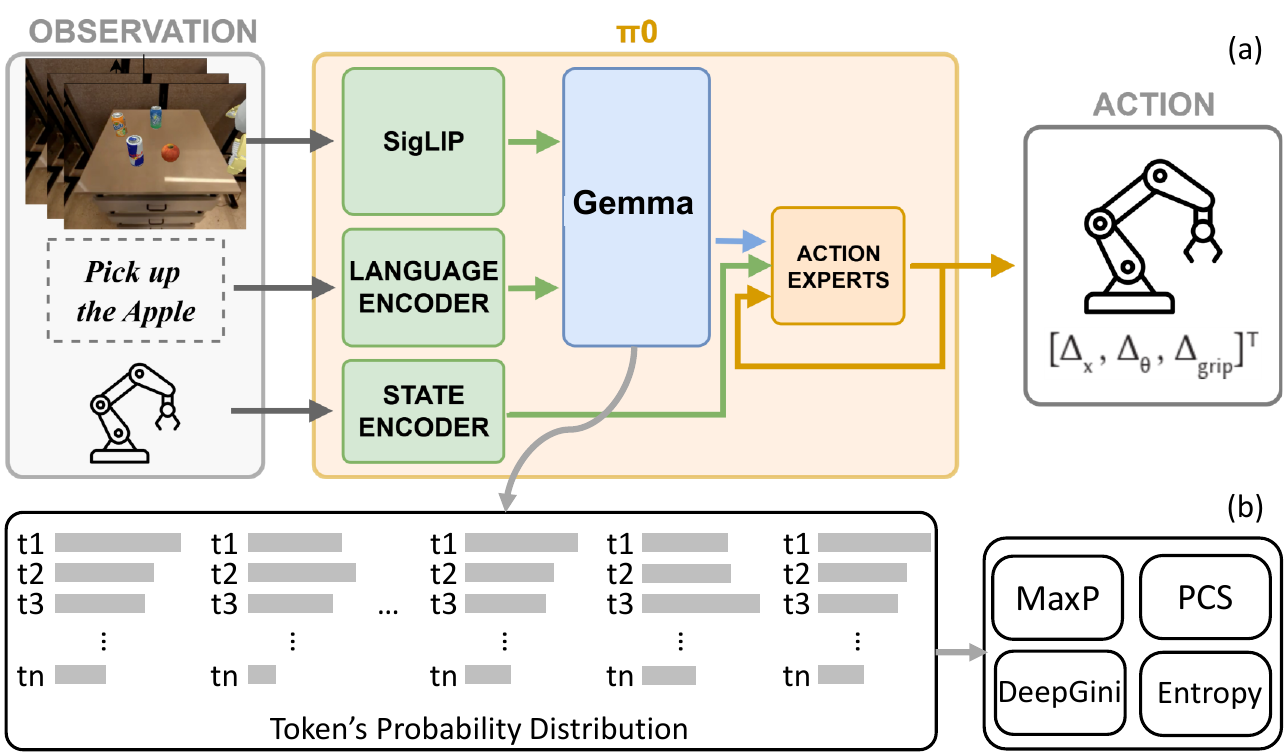}
    \caption{Architecture of \pimodel}
    \label{fig:Arch_PI0}
\end{figure}

$\pi_0$~\cite{black2024pi0visionlanguageactionflowmodel} is a 3.3B parameter VLA model trained with conditional flow matching~\cite{lipman2022flow,liu2022rectified, zhao2023learning} on a diverse mixture of datasets including the Open X-Embodiment dataset~\cite{o2024open}, the \revision{Bridge dataset}~\cite{walke2023bridgedata}, and DROID dataset~\cite{khazatsky2024droid} high-frequency dexterous data. Its architecture (See Figure~\ref{fig:Arch_PI0} (a)) consists of two main components: (1) a large VLM backbone built from SigLIP~\cite{zhai2023sigmoid} and Gemma-2B~\cite{beyer2024paligemma}, and (2) a specialized 300M-parameter action expert transformer~\cite{shazeer2017outrageously,lepikhin2020gshard} for modeling continuous actions. Robot observations, including multi-view RGB images, natural language prompts, and proprioceptive states, are tokenized and processed via the VLM. To model the distribution of future actions, \pimodel{} uses conditional flow matching, where each action expert functions as a denoising network, i.e., it reconstructs clean actions from noise-injected inputs, producing coherent and precise trajectories. Proprioceptive and action tokens are passed exclusively through the action expert, which operates under a bidirectional attention mask. At inference time, the action expert starts from random noise and progressively refines it into a coherent sequence of actions by denoising over several steps, guided by the learned flow field~\cite{liu2024playground}.

\begin{figure}[h]
    \centering
    
        \includegraphics[width=0.485\textwidth]{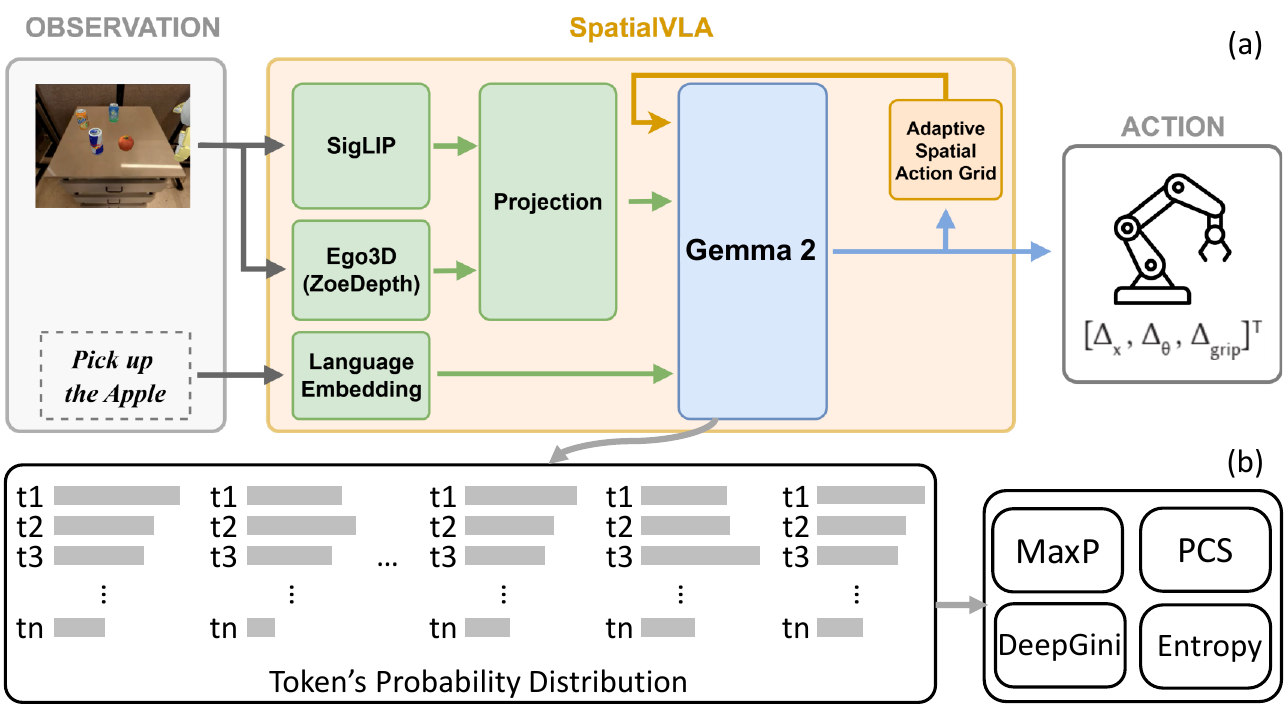}
    \caption{Architecture of SpatialVLA}
    \label{fig:Arch_SpatialVLA}
\end{figure}

\textbf{SpatialVLA}~\cite{qu2025spatialvla} is a 3B-parameter Vision-Language-Action (VLA) model pretrained on 1.1 million real-robot demonstrations from the Open X-Embodiment~\cite{o2024open} and RH20T~\cite{fang2024rh20t} datasets. The architecture of SpatialVLA consists of three main components (see Figure~\ref{fig:Arch_SpatialVLA} (a)): (1) a visual encoder that extracts features with Ego3D position encoding, (2) a spatial action tokenizer that uses adaptive action grids, and (3) a large language model as the backbone. In particular, SpatialVLA uses PaLIGemma-2 (Gemma 2)~\cite{steiner2024paligemma} as its backbone vision-language model. To extract visual features, SpatialVLA employs SigLIP~\cite{zhai2023sigmoid} as the 2D visual encoder, ensuring alignment between vision and language inputs. For Ego3D position features, it incorporates ZoeDepth~\cite{bhat2023zoedepth}, which estimates depth to derive the 3D position of each pixel. Finally, via Adaptive Action Grids, which discretize the continuous robot actions into adaptive spatial grids according to statistical action distributions on the whole robot episodes, the robot learns spatial action tokens on these grids to align cross-robot actions with the 3D spatial structure of the physical world. This eliminates the need for robot-camera extrinsic calibration and makes the model agnostic to specific robot setups, a key advantage for real-world deployment.

\subsection{Implementation of the token-based uncertainty metrics}

To calculate the token-based metrics, we operate at the individual token level instead of the final outputs of VLAs. Therefore, we need to identify the specific module within each VLA that generates token-level outputs. For \textbf{OpenVLA}, we take the token outputs from its LLM backbone, which is a 7B Llama2 language model (see Figure~\ref{fig:Arch_Openvla} (b)). The tokenizer used by Llama2 outputs 7 tokens, each of which can be detokenized as a control action. The generation of each token can be referred to as a classification problem, where the model selects one token with the highest probability from a token vocabulary of size 32,064 based on the predicted probability distribution over all possible tokens. Once the token probabilities are obtained, we compute the token-based uncertainty metrics as defined in Section~\ref{sec:UncertaintyMetrics}.

For $\mathbf{\pi_0}$, its token outputs are obtained from its PaliGemma-based backbone (see Figure~\ref{fig:Arch_PI0} (b)), consisting of the SigLIP vision encoder and the Gemma-2B language model. $\mathbf{\pi_0}$ utilizes the \textit{GemmaForCausalLM} model to load the Gemma-2B model. \textit{GemmaForCausalLM} is a pretrained model for causal language modeling that integrates the Gemma model with a language modeling head. The Gemma model encodes the input sequence and produces contextualized token representations, which are then transformed by the language modeling head into token-level predictions. The language modeling head outputs 5 tokens, each of which is selected from a vocabulary of size 257,152 based on the predicted token probability distribution. These token-level outputs are then used to compute the token-based metrics.

For \textbf{SpatialVLA}, similar to $\mathbf{\pi_0}$, its token outputs are obtained from its PaliGemma2-based backbone (see Figure~\ref{fig:Arch_SpatialVLA} (b)). PaliGemma2 shares the same model architecture as PaliGemma, but introduces several key improvements, including support for higher image resolutions and larger language model variants (e.g., Gemma-2B, 9B, and 27B). Therefore, we followed the same implementation as $\mathbf{\pi_0}$ to obtain the token probability distribution. The token outputs are obtained from Gemma2 model, which outputs 13 tokens, each of which is selected from a vocabulary of size 265,347. We then calculate the token-based metrics based on the token outputs.

\section{In depth analysis of the results}

\begin{figure*}[h]
    \centering
    \includegraphics[width=\textwidth]{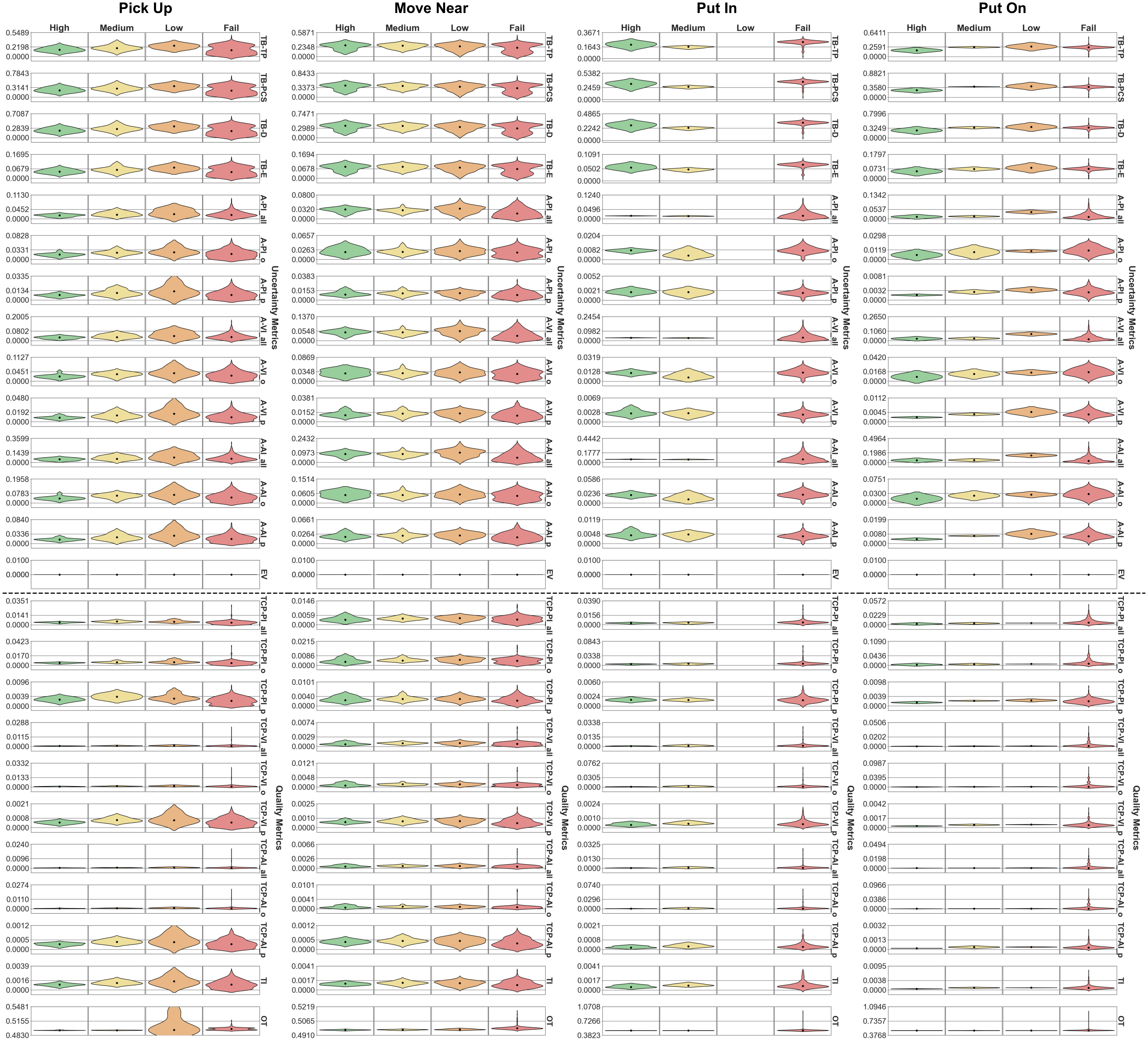}
    \caption{Distribution of average values across all metrics for each task type. The columns "High", "Medium", and "Low" correspond to the human evaluation of successfully completed tasks, while "Fail" shows the results for failed executions. Note that the y-axis range is independently scaled for each task and metric to better visualize score differences across quality levels. Results shown are for the OpenVLA model.}
    \label{fig:RQ2-openvla}
\end{figure*}

\begin{figure*}[h]
    \centering
    \includegraphics[width=\textwidth]{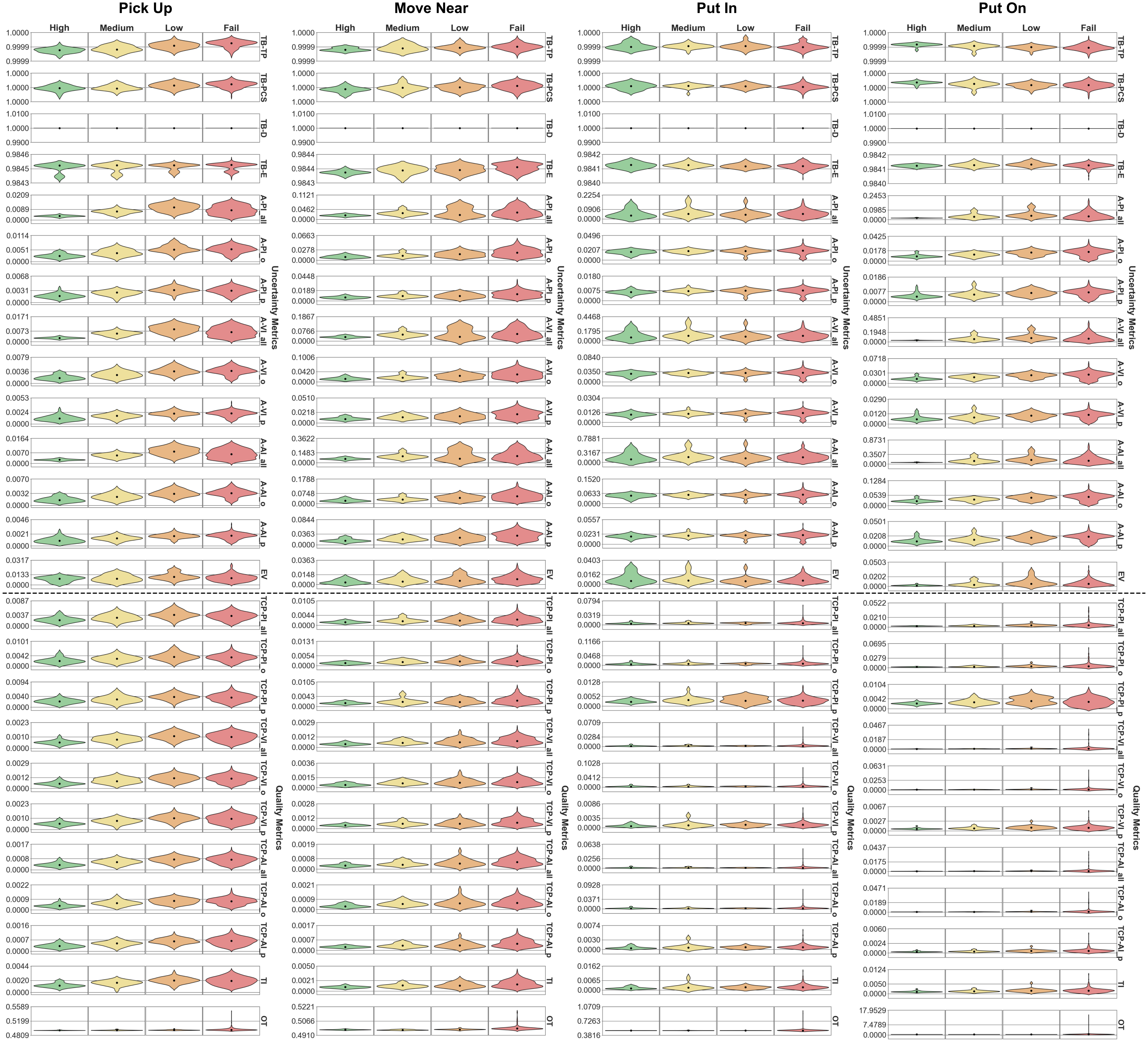}
    \caption{Distribution of average values for the \pimodel{} model}
    \label{fig:RQ2-pi0}
\end{figure*}

\begin{figure*}[h]
    \centering
    \includegraphics[width=\textwidth]{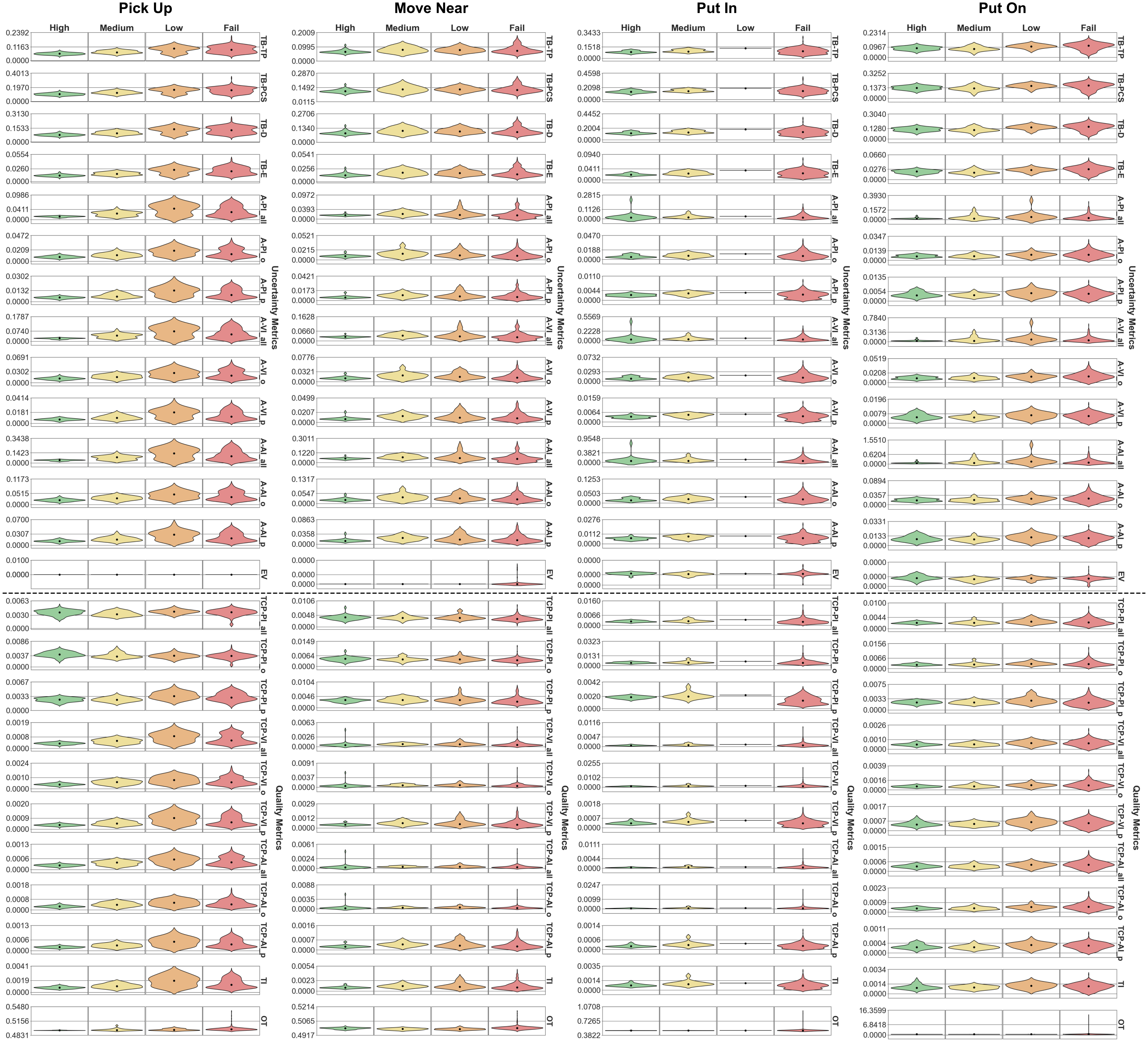}
    \caption{Distribution of average values for the SpatialVLA model}
    \label{fig:RQ2-spatialvla}
\end{figure*}

To support a deeper understanding of how the proposed uncertainty and quality metrics relate to actual task execution quality, Figure~\ref{fig:RQ2-openvla}, Figure~\ref{fig:RQ2-pi0} and Figure~\ref{fig:RQ2-spatialvla} present violin plots illustrating the distribution of the metric values for each human evaluation label. These plots serve as a visual complement to the quantitative results and help convey the variability and consistency of each metric across different test scenarios. Each violin plot represents the full distribution of metric values obtained from individual test cases, enabling a more detailed view of how well the metrics capture the trends in task-level performance. The numerical Spearman correlation values corresponding to these plots are provided in Table~\ref{tab:moveNearCorr}.

\end{document}